\pdfoutput=1
\documentclass[structabstract,natbib]{aa}  

\usepackage{natbib}
\usepackage{graphicx}
\usepackage{txfonts}
\begin{document}

\newcommand{\couno}{$^{12}$CO(1-0)}
\newcommand{\codue}{$^{12}$CO(2-1)}
\newcommand{\defhi}{{\it def}$_{\rm HI}$}
\newcommand{\micron}{$\mu$m}
\newcommand{\xco}{$X_{\rm CO}$}
\newcommand{\msolpc}{M$_{\odot}$ pc$^{-2}$}

   \title{The {\it Herschel}\thanks{{\it Herschel} is an ESA space observatory with science instruments provided by European-led Principal Investigator consortia and with important participation from NASA.} Virgo Cluster Survey}

   \subtitle{XI. Environmental effects on molecular gas and dust in spiral disks\thanks{Based on observations carried out with the IRAM 30m Telescope. IRAM is supported by INSU/CNRS (France), MPG (Germany) and IGN (Spain).}}

   \author{Ciro Pappalardo\inst{1}, Simone Bianchi\inst{1}, Edvige Corbelli\inst{1}, Carlo Giovanardi\inst{1}, Leslie Hunt\inst{1}, George J. Bendo\inst{6}, Alessandro Boselli\inst{4}, Luca Cortese\inst{5}, Laura Magrini\inst{1}, Stefano Zibetti\inst{1}, Sperello di Serego Alighieri\inst{1}, Jonathan Davies\inst{2}, Maarten Baes\inst{3}, Laure Ciesla\inst{4}, Marcel Clemens\inst{7}, Ilse De Looze\inst{3}, Jacopo Fritz\inst{3}, Marco Grossi\inst{8}, Michael Pohlen\inst{2}, Matthew W. L. Smith\inst{2}, Joris Verstappen\inst{3}, Catherine Vlahakis\inst{9}}

   \institute{
   Osservatorio Astrofisico di Arcetri - INAF, Largo E. Fermi 5, 50125 Firenze, Italy e-mail: {\tt cpappala@arcetri.astro.it}
   \and
   Department of Physics and Astronomy, Cardiff University, The Parade, Cardiff, CF24 3AA, UK
   \and
   Sterrenkundig Observatorium, Universiteit Gent, Krijgslaan 281 S9, B-9000 Gent, Belgium
   \and
   Laboratoire dÕAstrophysique de Marseille - LAM, Universit\'e d'Aix-Marseille \& CNRS, UMR7326, 38 rue F. Joliot-Curie, 13388 Marseille Cedex 13, France
   \and
   European Southern Observatory, Karl-Schwarzschild-Strasse 2, D-85748 Garching bei Munchen, Germany 
   \and
   Jodrell Bank Centre for Astrophysics, Alan Turing Building, School of Physics and Astronomy, University of Manchester, Manchester M13 9PL 
   \and
   Osservatorio Astronomico di Padova, Vicolo dell'Osservatorio 5, 35122 Padova, Italy
   \and
   CAAUL, Observatorio Astronomico de Lisboa, Universidade de Lisboa, Tapada de Ajuda, 1349-018, Lisboa, Portugal 
\and
Joint ALMA Office, Alonso de Cordova 3107, Vitacura, Santiago, Chile / European Southern Observatory, Alonso de Cordova 3107, Vitacura, Casilla 19001, Santiago 19, Chile
             }

   \date{}
  \abstract
   {} 
   {We investigate the dust-to-gas mass ratio and the environmental effects on the various components of the interstellar medium for a spatially resolved sample of Virgo spirals.}
   {We have used the IRAM-30m telescope to map over their full extent NGC 4189, NGC 4298, NGC 4388, and NGC 4299 in the \couno\ and the \codue\ lines. We observed the same lines in selected regions of NGC 4351, NGC 4294, and NGC 4424. The CO observations are combined with Herschel maps in 5 bands between 100-500 $\mu$m from the HeViCS survey, and with HI data from the VIVA survey, to obtain spatially resolved dust and gas distributions. We studied the environmental dependencies by adding to our sample eight galaxies with \couno\ maps from the literature.}
   {We estimate the integrated mass of molecular hydrogen for the galaxies observed in the CO lines. We find molecular-to-total gas mass fractions between 0.04 $\le f_{mol} \le$ 0.65, with the lowest values for the dimmest galaxy in the B-band. The integrated dust-to-gas ratio ranges between 0.011 and 0.004. For the 12 mapped galaxies we derive the radial distributions of the atomic gas, molecular gas, and dust. We also study the effect of different CO-to-H$_2$ conversion factors. Both the molecular gas and the dust distributions show steeper radial profiles for HI-deficient galaxies and the average dust-to-gas ratio for these galaxies increases or stays radially constant. On scales of $\sim$ 3 kpc, we find a strong correlation between the molecular gas and the 250 $\mu$m surface brightness that is tighter than average for non-deficient galaxies. The correlation becomes linear if we consider the total gas surface mass density. However, the inclusion of atomic hydrogen does not improve the statistical significance of the correlation.} 
   {The environment can modify the distributions of molecules and dust within a galaxy, although these components are more tightly bound than the atomic gas.}
   

\titlerunning{Environmental effects on molecular gas and dust in Virgo spirals}
\authorrunning{Pappalardo et al.}
   \maketitle

\section{Introduction}

The stellar and gas distributions in a cluster galaxy can be drastically modified through environmental effects. In Virgo, a young, close ($d \sim$ 17 Mpc) and still dynamically active cluster \citep[see][and references therein]{bos}, the environment effects are clearly seen for the atomic gas. Through tidal interaction and/or ram pressure stripping a spiral galaxy can lose a considerable fraction of its neutral gas and become deficient in HI \citep{dav3,gio,hay,cay,chu}.

For the molecular gas-phase the situation is more complex. The typical density of a molecular cloud core is 10$^5$ cm$^{-3}$ \citep{case}, and the molecular gas is more tightly bound in the galaxy potential, being confined to the inner part of the disk which may prevent gas removal via stripping. According to \citet{ken2, ken3}, Virgo spirals show no clear signs of deficiency in the molecular component. In the Coma cluster, where the ram pressure is expected to be stronger due to the high ICM densities, there is no clear evidence of molecular deficiency \citep{cas}. However, as noted by \citet{bos}, the sample of \citet{cas} is selected according to IR luminosity, and their results are biased toward strong CO emitters \citep[see also][]{bos3, bos4}. Using the same data of \citet{ken3}, \citet{ren} found evidence of a CO deficit in some Virgo spiral, and \citet{vol2} detected in NGC 4522 molecular gas displaced from the galactic disk because of ram pressure. 

\citet{bos2}, \citet{fum} and \citet{fum2} found a molecular depletion in highly HI deficient galaxies, where the HI is depleted well inside the optical disk. Since not all the HI-deficient galaxies examined in their papers are cluster galaxies, it is of interest to analyze a sample of normal and HI-deficient galaxies in a cluster. Also, it is still not clear if the depletion of the molecular component in HI-deficient galaxies is due to ram pressure stripping of the molecular component, or to the lack of HI from which molecular clouds form.

The {\it Herschel} Virgo Cluster Survey \citep[HeViCS;][]{dav,aul} has mapped an area of 84 square deg$^2$ in the Virgo cluster, and one of the goals of the project is to investigate mechanisms of gas recycling in clusters of galaxies. \citet{cor2} analyzed the galaxies in the {\it Herschel} Reference Survey \citep[HRS;][]{bos5}, including many Virgo galaxies observed within HeViCS, and found that the global dust-to-stellar mass ratio decreases as a function of HI deficiency, providing strong evidence of dust stripping. Furthermore they found that at fixed stellar mass, the dust-to-HI mass ratio decreases with the HI deficiency, with a correlation significantly weaker than for the dust-to-stellar mass ratio. They argue that this mild correlation is a consequence of the greater extension of the HI disks which produces a rapid decline of the dust-to-HI ratio in the outer regions of galaxies. 
For a sample of 35 HeViCS Virgo spirals with available integrated CO fluxes, \citet{corb2} confirmed that the dust-to-stellar mass ratio decreases and the dust-to-total gas ratio increases when the galaxies are more HI-deficient.

The investigation of the internal properties of galaxies can help understand environmental effects. \citet{cor} investigated 15 Virgo spiral galaxies, and found that the radial distributions of dust in HI-deficient galaxies are truncated, already in the inner regions of the disks. 

It is therefore important to study the relation between the radial distribution of dust and the different gas components within a spiral disk. To reach this goal, it is necessary to have a sample of galaxies observed in all the interstellar medium (ISM) components. The proximity of Virgo allows a good resolution of the gas and dust components (a typical beam size of 20$''$ corresponds to a resolution of 1.6 kpc). Several Virgo spiral galaxies have been mapped with the Very Large Array (VLA) in HI, within the survey "VLA Imaging of Virgo spirals in Atomic gas" \citep[VIVA;][]{chu}. About half of them are mapped in the FIR within the HeViCS project \citep{dav}. However, only a subset of these have observations that trace the spatial distribution of the molecular component, major axis profiles or maps such as those shown in \citet{chu2}, \citet{kun}. To complete the sample we thus began a project to map the CO emission in the remaining low luminosity objects.

In this paper we present \couno\ and \codue\ emission-line observations for seven galaxies, obtained at the IRAM-30m telescope. We use these new data to investigate the balance between different gas phases in spiral disks and their relation with the dust component as a function of the environment. The paper is organized as follows: in Sec.~\ref{galsample}, we describe the sample, and the ancillary data used in the paper. In Sec.~\ref{galdata} we present the new IRAM-30m observations and the relative results. In Sec.~\ref{disk} we discuss in detail the radial distribution of the dust and gas phases for the galaxies mapped in our new campaign and for additional Virgo spirals with available CO, HI, and dust maps. We also investigate the variations of the dust-to-gas ratio across the disk. In Sec.~\ref{pixpix} we perform a pixel-by-pixel analysis of the relation between the far-infrared (FIR) emission and the gas surface mass density. The main conclusions are summarized in Sec.~\ref{conc}.

\section{The sample and ancillary data}
\label{galsample}

We have compiled a sample of bright ($m_B \le$ 13) late-type galaxies in the Virgo cluster with available VIVA and HeViCS observations. Here we present \couno\ and \codue\ emission-line maps of four spirals (NGC\,4189, NGC\,4298, NGC\,4299, NGC\,4388) obtained with the IRAM-30m telescope. We also observed a few selected regions of 3 additional low luminosity galaxies, NGC\,4351, NGC\,4294, and NGC\,4424. To enlarge the sample of fully mapped galaxies we also consider eight additional galaxies mapped in \couno \ at the Nobeyama 45 m telescope by \citet{kun} at comparable spatial resolution. This sample of 12 mapped galaxies spans a wide range of morphological types, from Sa (NGC\,4424) to Scd (NGC\,4299). Since we are interested in studying the environmental effects on the gas distribution, we excluded from the sample gravitationally interacting pairs.

Interferometric observations of HI for the VIVA sample are described in \citet{chu}. The velocity resolution of VIVA (10 km s$^{-1}$) is comparable to the final resolution of our CO data (10.4 km s$^{-1}$). The sensitivity of the VIVA observations is around 3-5 $\times$ 10$^{19}$ HI atoms cm$^{-2}$ (3$\sigma$, 0.3 \msolpc) at a typical angular resolution of $\sim$25$''$. We quantify the HI deficiency, \defhi, in our sample using the prescription of \citet{chu}, which assumes a mean HI mass-diameter relation, independently of morphological type. Our galaxies span a wide range of values, -0.43 $\le$ \defhi \ $\le$ 1.47, revealing different stages of interaction with the environment. Even if we adopt an HI deficiency criterion based on morphological type \citep[e.g.][]{gio}, our sample still covers a wide range of \defhi: from -0.2 to 1.1. Given that the uncertainties in \defhi \ are $\sim$ 0.25-0.3 dex we shall consider a galaxy as non-deficient only if it has \defhi \ $\lesssim 0.4$.

FIR images for dust emission at 100, 160, 250, 350, and 500 $\mu$m come from the HeViCS full depth data set \citep[see][ for details on the data reduction]{aul}. The beam Full Width Half Maximum for these wavelengths are 9$''$, 13$''$, 18\farcs2, 24\farcs5, and 36$''$, respectively. 

The dust surface mass density is derived as in \citet{mag} \citep[see also][]{smi, ben}. The five images in the {\it Herschel} photometric bands were convolved to the poorer resolution of the 500 $\mu$m, then registered on the same pixel grid. Each pixel was fitted with a single temperature modified black body, with dust emissivity $\kappa \propto \nu^\beta$ and $\beta$ = 2; the filter response function was taken into account for the color correction. As shown by \citet{dav2} for global fluxes, and by \citet{mag} for pixel-by-pixel surface brightness analysis, our choice of the dust emissivity provides a satisfactory fit to the spectral energy distribution. If we exclude the notorious uncertainty of the dust emissivity, the uncertainties due to the fitting procedure amounts to 10-15\% for the full depth HeViCS data set. Other sources of uncertainty in the determination of the total mass are discussed in \citet{mag}. 

In Table \ref{sample} we summarize the main properties of the galaxies in the sample, and in Appendix \ref{single} we describe more in detail the galaxies for which we present new CO observations.

\begin{table*}
\caption{General properties of the sample}
\begin{center}
\begin{tabular}{c c c c c c c c c c c c c c}
\hline \hline
Galaxy & RA  & Dec & Type & m$_B$ & D & V$_{hel}$ & a & i & P.A. & \defhi & $Z_c$\\
 &deg& deg & & & Mpc & km s$^{-1}$ & $'$ & deg &deg&& \\
 (1) & (2) & (3) & (4) & (5) & (6) & (7) & (8) & (9) & (10) & (11) & (12)\\
 \hline
\multicolumn{12}{c}{Targets of our observations}\\
\\
 NGC     4189 &        183.447 & 13.42 & Sc & 12.69 & 32 &     2114 & 2.40 &       45 &       66 & 0.25 & 9.09  \\ 
 NGC     4294 &        185.324 & 11.51 & Sc & 12.68 & 17 &      355 & 3.24 &       70 &      155 & -0.11 & 8.95  \\ 
 NGC     4298 &        185.386 & 14.61 & Sc & 11.95 & 17 &     1136 & 3.24 &       57 &      140 & 0.41 & 9.08  \\ 
 NGC     4299 &        185.419 & 11.50 & Scd & 12.99 & 17 &      232 & 1.74 &       22 &       42 & -0.43 & 8.83  \\ 
 NGC     4351 &        186.006 & 12.20 & Sc & 12.99 & 17 &     2316 & 1.20 &       49 &       61 & 0.23 & 8.88  \\ 
 NGC     4388 &        186.445 & 12.66 & Sab & 11.87 & 17 &     2515 & 5.62 &       75$^a$ &       92 & 1.16 & 9.09  \\ 
 NGC     4424 &        186.798 & 9.42 & Sa & 12.33 & 23 &      437 & 3.63 &       62 &       95 & 0.97 & 9.08  \\ 
 \hline
\multicolumn{12}{c}{Galaxies with \couno \ maps from \citet{kun}}\\
\\
 NGC     4192 &        183.451 & 14.90 & Sb & 10.73 & 17 &     -139 & 9.77 &       78 &      155 & 0.51 & 9.08  \\ 
 NGC     4254 &        184.707 & 14.42 & Sc & 10.45 & 17 &     2404 & 5.37 &       30 &       68 & -0.10 & 9.09  \\ 
 NGC     4321 &        185.728 & 15.82 & Sc & 10.02 & 17 &     1575 & 7.41 &       33 &       30 & 0.35 & 9.07  \\ 
 NGC     4402 &        186.532 & 13.11 & Sc & 12.64 & 17 &      230 & 3.89 &       78 &       90 & 0.74 & 9.08  \\ 
 NGC     4501 &        187.996 & 14.42 & Sbc & 10.50 & 17 &     2282 & 6.92 &       59 &      140 & 0.58 & 9.05  \\ 
 NGC     4535 &        188.585 & 8.20 & Sc & 10.73 & 17 &     1964 & 7.08 &       46 &        0 & 0.41 & 9.09  \\ 
 NGC     4569 &        189.208 & 13.16 & Sab & 10.08 & 17 &     -221 & 9.55 &       65 &       23 & 1.47 & 9.08  \\ 
 NGC     4579 &        189.431 & 11.82 & Sab & 10.52 & 17 &     1516 & 5.89 &       38 &       95 & 0.95 & 9.06  \\

\hline \hline\end{tabular}\end{center}
Column 1: name; col. 2-3: equatorial coordinate (J2000); col. 4: morphological type \citep{bin,bin2}; col. 5: B-band magnitude; col. 6: distance; col. 7: heliocentric velocity; col. 8: major axis optical diameter; col. 9: inclination; col. 10: Position Angle; col. 11: HI deficiency parameter; col. 12: central oxygen abundance expressed as 12+log O/H from the mass-metallicity relation of \citet{tre}. Data sources: col. 1-7, GOLDMine \citep{gav2}; col. 8-11, \citet{chu}; col. 12, \citet{corb2}.

$^{a}$ The inclination of this galaxy has been computed directly from the axis ratio.
\label{sample}
\end{table*}

\section{IRAM 30-m observations}
\label{galdata}
 
The CO observations were carried out at the IRAM 30-meter millimeter-wave telescope on Pico Veleta (Spain) during two observing runs in June 2010 and April 2011.
 
 \subsection{Observational setup and data reduction}
 
We observed the $^{12}$CO(1-0) and $^{12}$CO(2-1) rotational transitions, with rest frequencies of 115.271 and 230.538 GHz, respectively. As front-end we used the Eight MIxer Receiver \citep[EMIR;][]{car}, and adopted three different back-ends:
\begin{itemize}
\item the 4MHz filterbank, used for the \couno \ line only, providing a total band width of 4GHz, and a channel width of 4MHz, i.e. 10.4 km s$^{-1}$;
\item the Wideband Line Multiple Autocorrelator (WILMA) with a bandwidth of 1860 channels (4GHz), and a spectral resolution of 2MHz. This corresponds to a spectral resolution of 5.2, and 2.25 km s$^{-1}$ for the \couno \ and \codue \ lines;
\item the VErsatile SPectrometer Assembly (VESPA) for which we selected the resolution of 1.25MHz, with a smaller bandwidth, 420 MHz. This setup gives a spectral resolution of 3.25, and 1.62 km s$^{-1}$ for the \couno \ and \codue \ lines;
\end{itemize}

We performed a flux calibration every 15 minutes, checked the pointing every 2h, and the focus every 6h, as customary. The spectra were calibrated in main-beam temperature, assuming the main-beam and forward efficiencies $\eta_{mb}$ = 0.78 and $\eta_{for}$ = 0.95 for the $^{12}$CO(1-0) and  $\eta_{mb}$ = 0.59, $\eta_{for}$ = 0.91 for $^{12}$CO(2-1), according to the values given in the IRAM-30m website as of July 2011. The main-beam temperature is T$_{mb} = \eta_{for}/\eta_{mb} \cdot$ T$_A^*$, where T$_A^*$ is the antenna temperature \citep{wil}.

In order to increase the S/N ratio we averaged the spectra taken with all the backends to a common resolution of 10.4 km s$^{-1}$. We removed the baseline outside the signal window by fitting polynomial functions\footnote{Most of data reduction was performed using GILDAS software: \it{http://www.iram.fr/IRAMFR/GILDAS}} of orders between 1 and 3. The half power beam widths (HPBW) are 22\farcs5 for $^{12}$CO(1-0) and 11\farcs25 for $^{12}$CO(2-1). We observed our sample in 2 different observing modes: On The Fly (OTF) map and Position Switching (PS). The choice of the observing mode was based on the estimated total CO brightness from FIR emission.

\begin{table*}
\caption{Integrated quantities for observations in OTF mode}
\begin{center}
\begin{tabular}{c c c c c c c c cccccc}
\hline \hline
\\
Galaxy & Map Size  & $\sigma_{1-0}$ & $\sigma_{2-1}$ & $\overline{I_{1-0}}$& $\overline{I_{2-1}}$  & $\overline{I_{2-1}/I_{1-0}}$ &$\overline{\Sigma^{H_2}_{1-0}}$ & M$^{H_2}_{1-0}$\\
	   &   $'' \ \times \ ''$	& mK & mK   & K km s$^{-1}$       & K km s$^{-1}$  	&	&	M$_\odot$ pc$^{-2}$ 	& $10^8 \ M_{\odot}$ \\
	   (1 ) & (2) & (3) & (4) & (5) & (6) & (7) & (8) & (9)\\
 \hline
 &&&&&&&&\\
 NGC 4189 & 96 $\times$ 75 & 7 & 10 & 3.73$\pm$0.34 & 3.43$\pm$0.78 &  0.94	& 11.95$\pm$1.1 & 17.4$\pm$1.3\\
 NGC 4298 & 120 $\times$ 60& 11 & 14 & 19$\pm$0.66 & 6.04$\pm$2.4  &0.76 & 23.32$\pm$2.1 &10.4$\pm$0.8\\
 NGC 4299$^*$ & 104$\times$104 & 19 & 42 & 1.28$\pm$0.28 & 1.2$\pm$0.7 &  .. & 4.1$\pm$2.6 & 1.1$\pm$0.2\\
NGC 4388 &117$\times$ 42 & 13 & 39 & 8.93$\pm$0.75 &  10.8$\pm$2.6  & 0.9 &28.55$\pm$2.4 & 7.13$\pm$0.6\\
\hline \hline\end{tabular}\end{center}
Column 1: name; col. 2: map size; col. 3, 4: measured uncertainties in 10.4 km s$^{-1}$ channels for a 22\farcs5 beam; col. 5, 6: average value of the integrated intensity; col. 7: average $^{12}$CO(2-1)/$^{12}$CO(1-0) intensity ratio; col. 8: average surface mass density of H$_2$ using \couno\ and the Galactic \xco; col. 9: total H$_2$ mass.

*For this galaxy, because of the low S/N ratio, the results shown were obtained integrating inside an aperture of $R$ = 35$''$ (see Sec.~\ref{galmap} for details).
\label{mapsize}
\end{table*}
\begin{table*}
\caption{Integrated quantities for observations in PS mode}
\begin{center}
\begin{tabular}{c c c c c c c c c c c c}
\hline \hline
\\
Galaxy & $\Delta$RA, $\Delta$Dec &  $\sigma_{1-0}$ & $\sigma_{2-1}$ & I$_{1-0}$ &  I$_{2-1}$& $I_{2-1}/I_{1-0}$  & $\Sigma^{H_2}_{1-0}$   & M$_{H_2}$\\
&$''$& mK & mK & K km s$^{-1}$ & K km s$^{-1}$ &&  M$_\odot$ pc$^{-2}$ & 10$^8$ M$_{\odot}$\\
(1) &(2)&(3)&(4)&(5)&(6)&(7)&(8)&(9)\\
 \hline
 \\
NGC 4351 & +0.0,+0.0 & 4 & 13 &2.29$\pm$0.14 & 2.3$\pm$0.31 & 1.00 &7.34$\pm$0.44  & 0.71\\
NGC 4294 & +0.0,+0.0 & 4 & 4 & 2.66$\pm$0.11 & 1.36$\pm$0.12 & 0.51 & 8.5$\pm$0.35    & 0.73\\
	    & +9.3,-19.9 & 4 & 8 & 1.08$\pm$0.18 & 0.99$\pm$0.23 & 0.92 &3.47$\pm$0.58\\
NGC 4424 & +0.0,+0.0 & 6 & 10 & 9.41$\pm$0.24 & 9.0$\pm$0.4 & 0.96&30.1$\pm$0.77   & 4.2\\
	    & +21.9,-1.9 & 3 & 13 & 2.28$\pm$0.16 & 2.7$\pm$0.5 &1.18& 7.31$\pm$0.52 \\
	    & +43.8,-3.8 & 11 & 11 & 1.08$\pm$0.25 & 1.02$\pm$0.37 & 0.94& 3.67$\pm$0.8 \\
	    & -21.9,+1.9& 11 & 13 & 6.31$\pm$0.4 & 5.7$\pm$0.49 &0.9& 20.2$\pm$1.28 \\
	    & -43.8,+3.8& 11 & 11 & 1.05$\pm$0.16 & 1.04$\pm$0.5 & 0.99 & 3.35$\pm$0.52 \\		
\hline \hline\end{tabular}\end{center}
Column 1: galaxy name; col. 2: observed offset with respect to the galaxy center; col. 3, 4: 1$\sigma$ rms associated to \couno\ and \codue\ lines, respectively; col. 5, 6: integrated CO flux ($\int T d$v) for $^{12}$CO(1-0) and $^{12}$CO(2-1); col. 7: $^{12}$CO(2-1)/ $^{12}$CO(1-0) intensity ratio; col. 8: surface mass density derived from the \couno\ line using the Galactic \xco; col. 10: total H$_2$ mass (see Sec.~\ref{galmap} for details).
\label{integ}
\end{table*}

In OTF mode the telescope moves smoothly across the target and collects spectra while scanning. We set the integration time per position to 2 sec and the distance between each position along the scan to 4$''$, below the critical Nyquist frequency. After each scan a reference spectrum is taken in a region with no sources. Several scans along a chosen direction were taken to cover an area of about 1 square arcmin. The observations were repeated several time along parallel and perpendicular scan directions to reach the required sensitivity. In total, we observed NGC 4189 for 8.1h, NGC 4298 for 4h, NGC 4388 for 2.4h, and NGC 4299 for 4h. The sensitivities we achieved for the \couno \ and \codue \ lines are shown in columns 3, 4 of Table \ref{mapsize}. The size of each map is given in column 2 of Table \ref{mapsize}. Scans were done along the RA and DEC directions, with the exception of NGC 4298, where a scan direction parallel and perpendicular to the major axis was used.

The map pixel size is chosen in order to achieve roughly Nyquist sampling, 9\farcs3 for $^{12}$CO(1-0) and 4\farcs7 for $^{12}$CO(2-1) (HPBW/2.4). Spectra at each pixel are summed using the sigma-weighted average to obtain the map. The observations have been smoothed with a gaussian of size 1/3 of the telescope beam to gain in sensitivity at the expense of a slightly larger beam size \citep[see][]{man}. In the following analysis, the $^{12}$CO(2-1) maps have been convolved with an appropriate bidimensional Gaussian kernel to match the resolution of the $^{12}$CO(1-0). 

To estimate the Moment 0 and Moment 1 maps we defined in each spatial pixel a spectral signal window and integrated within it. We calculated the noise $\sigma$ outside the window and then estimated the noise related to the signal as N = $\sigma \cdot \Delta $v /$\sqrt{N_{ch}}$, where $\Delta$v is the width of the window in km s$^{-1}$ and $N_{ch}$ is the number of channels inside the signal window. In the final cube we took into account only the pixels with S/N $\ge$ 3. 

\smallskip

Three galaxies with expected fainter CO emission were observed in PS mode at a few selected positions, with the telescope switching between the galaxy and an offset position for background subtraction. In these observations we integrated for 15 minutes on each position and achieved the sensitivities shown in col. 3 and 4 in Table \ref{integ}. The original plan was to observe several positions along the major axis spaced by the \couno\ beamsize. However, due to bad weather conditions and time limitations this was possible only for NGC 4424, which was observed in 5 positions. For NGC 4294 it was possible to observe only the galaxy center and another position, and finally for NGC 4351 we only observed the galaxy center. The observed positions with respect to the center are shown in col. 2 of Table \ref{integ}.
\subsection{Results on individual galaxies}
\label{galmap}

Table \ref{mapsize} shows the average value of the integrated intensity for \couno\ and \codue\ lines (column 5 and 6), together with the average intensity line ratio (column 7) for galaxies observed in OTF mode. In column 8 and 9 we report the average surface brightness and the total mass of molecular hydrogen, obtained integrating over the pixels with surface mass densities above 3$\sigma$. We converted the \couno\ integrated intensities into molecular gas masses using the standard Galactic conversion factor \xco\ = $N_{H_2}/I_{CO} = 2 \cdot 10^{20}$ cm$^{-2}$ [K km s$^{-1}$]$^{-1}$, estimated in the solar neighborhood \citep{str,dam}. In Sec. \ref{disk} we investigate how the choice of different \xco \ factors affects the estimate of molecular hydrogen surface mass densities. We estimate the molecular fraction, defined as $f_{mol} =  M_{H_2}/(M_{H_2} + M_{HI})$, using the mass of atomic hydrogen given by \citet{chu}.

\begin{figure*}
\begin{center}
\includegraphics[clip=,width= .42\textwidth]{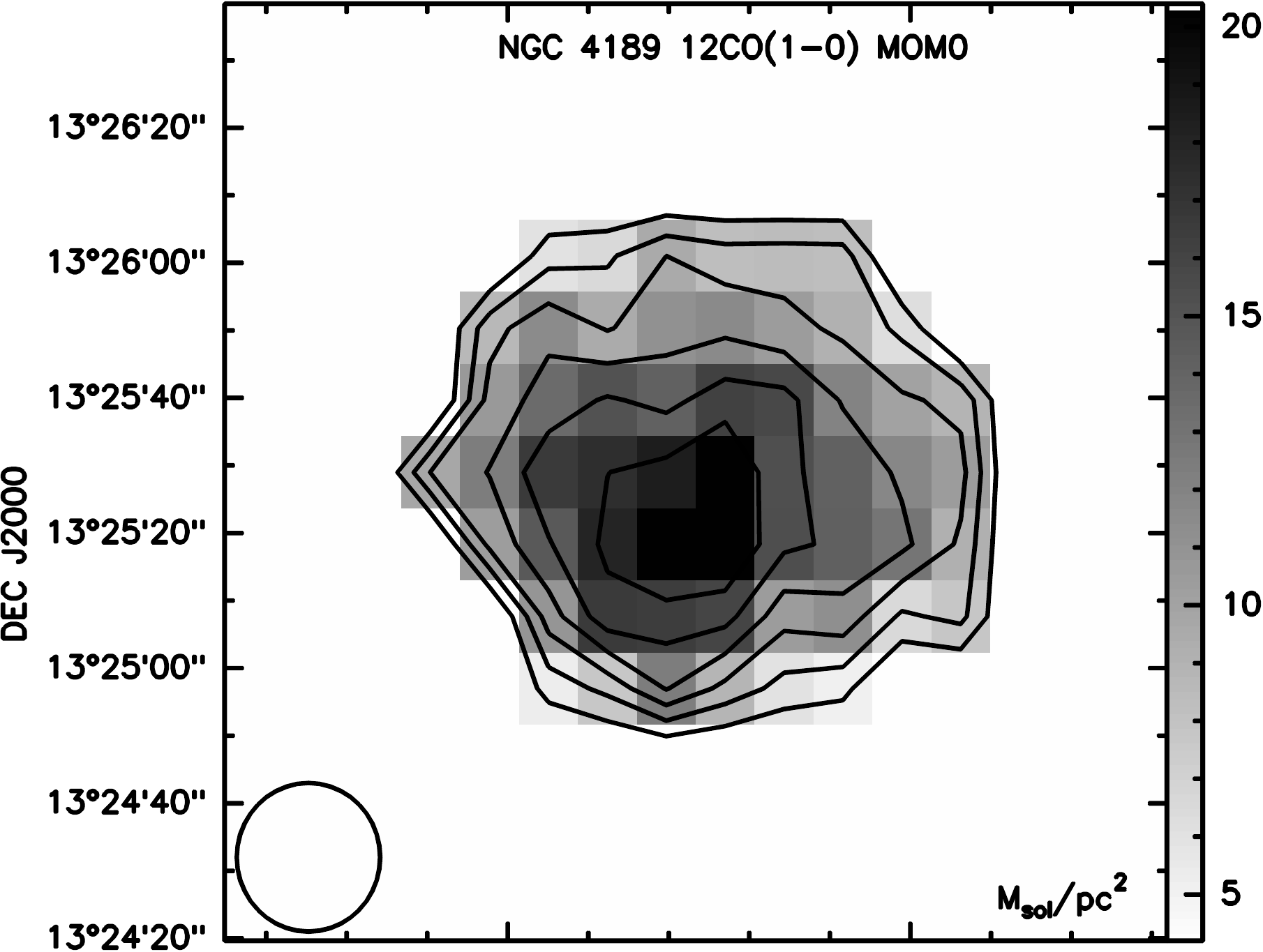}
\includegraphics[clip=,width= .35\textwidth]{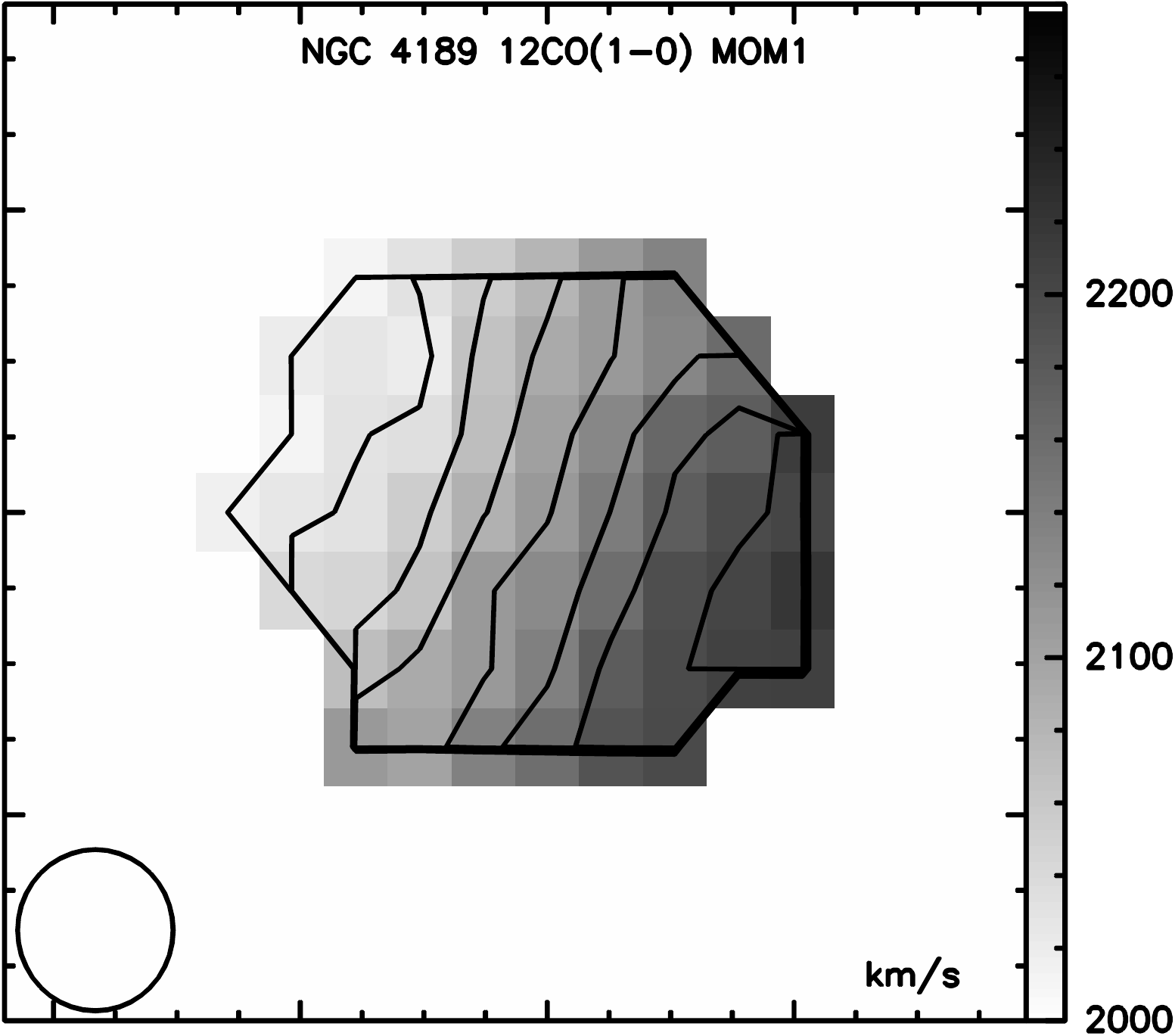}
\includegraphics[clip=,width= .41\textwidth]{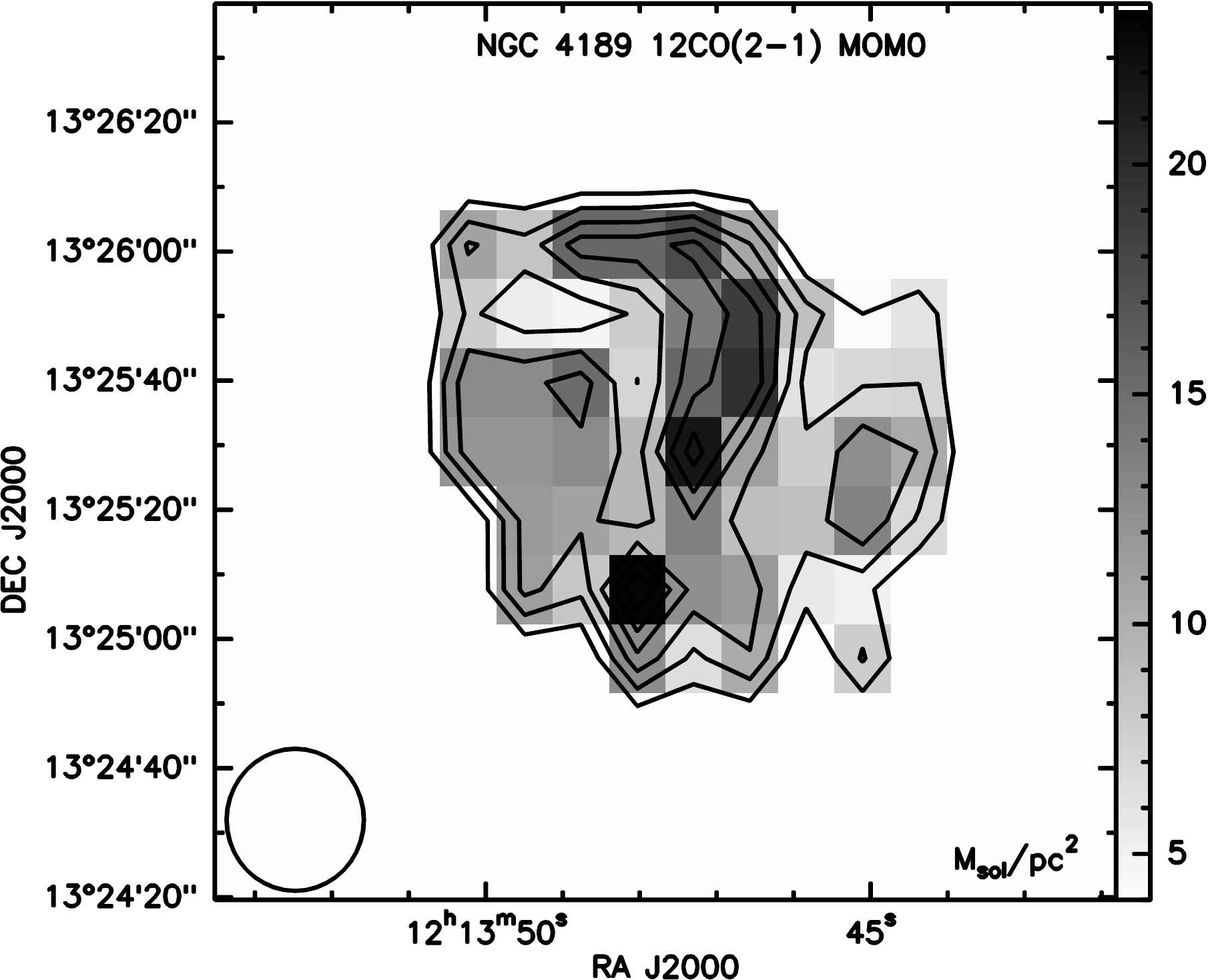}
\includegraphics[clip=,width= .34\textwidth]{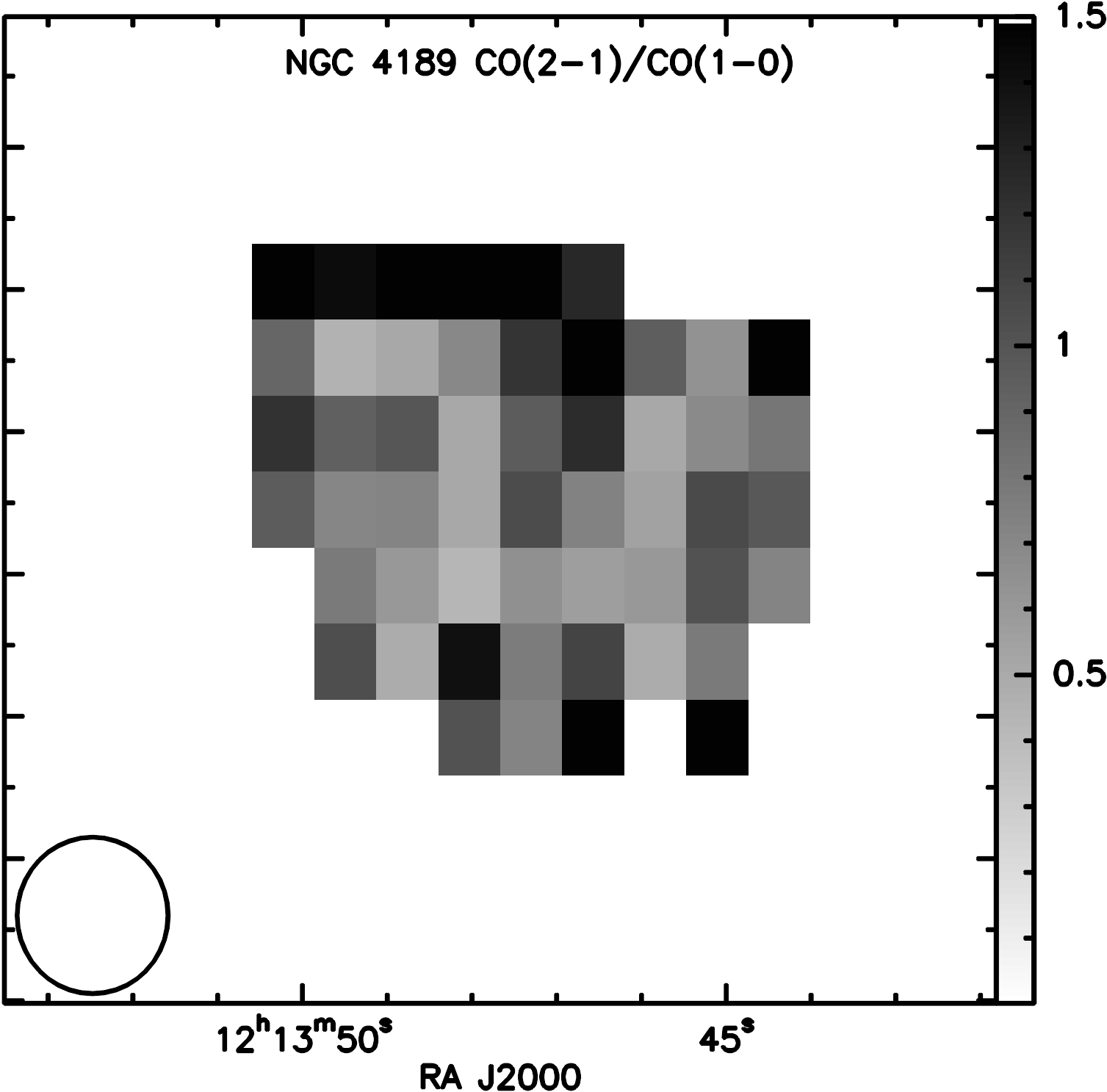}
\includegraphics[clip=,width= .65\textwidth]{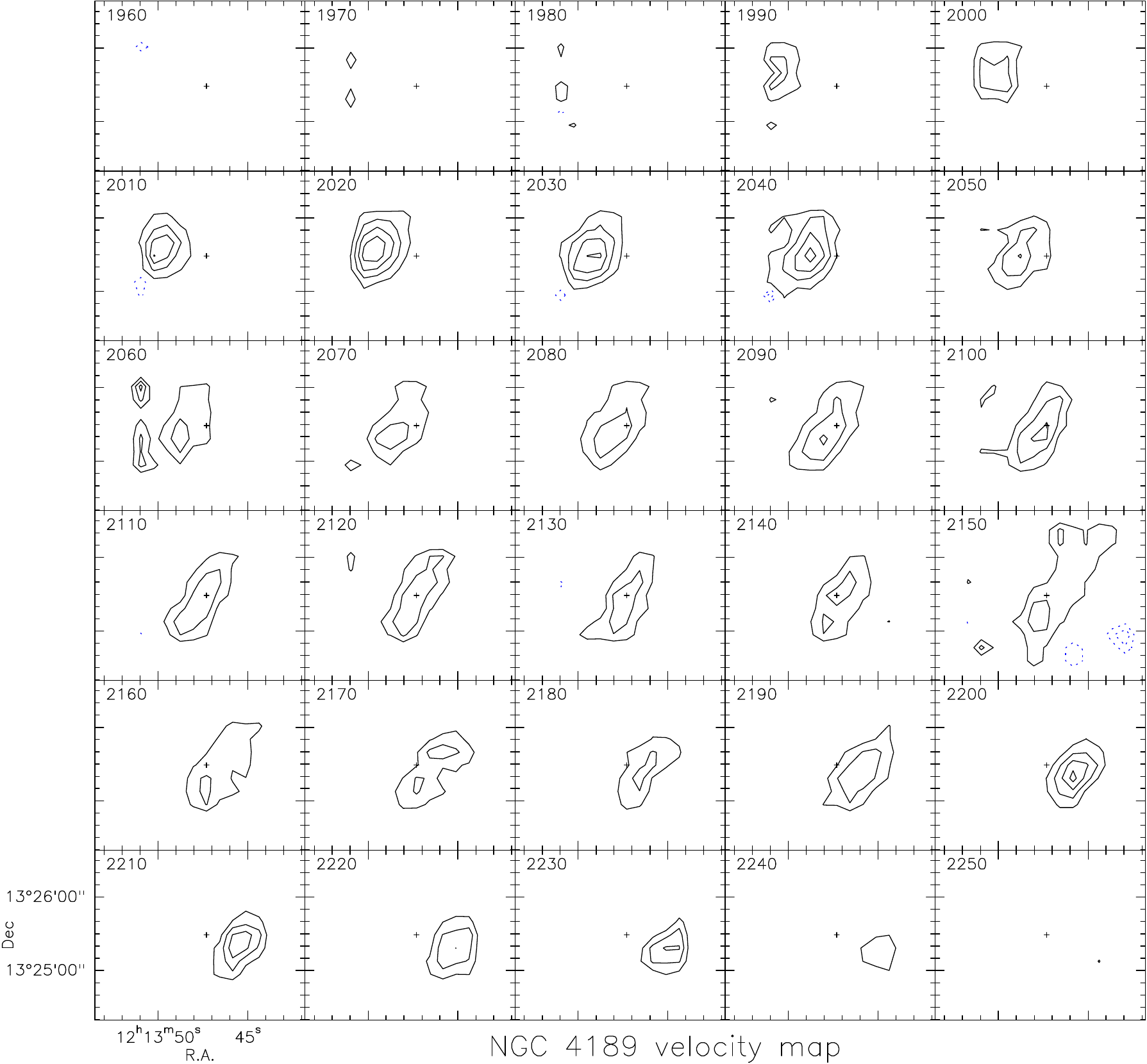}\end{center}
\caption{NGC 4189 maps. Top left: $^{12}$CO(1-0) Moment 0 map, contours: 4.2, 6.9, 9.6, 12.3, 15, 17.7 M$_{\odot}$ pc$^{-2}$. Top right: $^{12}$CO(1-0) Moment 1 with 2000 $\le$ v $\le$ 2230 and $\Delta$v = 28 km s$^{-1}$. Middle left: $^{12}$CO(2-1) Moment 0 map, contours: 4, 7.25, 10.5, 13.75, 17, 20.25, 23.5 M$_{\odot}$ pc$^{-2}$. Middle right: $^{12}$CO(2-1)/$^{12}$CO(1-0) ratio. The beam size is illustrated in the bottom left of each panel. Bottom panel: channels map of \couno\ with 1980 $\le$ v $\le$ 2250 km s$^{-1}$ and $\Delta$v = 10 km s$^{-1}$.}
\label{map4189}
\end{figure*}

In Figs. \ref{map4189}, \ref{map4298}, \ref{map4388} the CO maps obtained in OTF mode are presented in the following order: in the top left and right panels we show the moment 0 and moment 1 map of the \couno\ intensity emission; in the middle left and right panels the moment 0 \codue\ map and the \codue/\couno\ line intensity ratio are shown. The channel velocity maps with $\Delta$v = 10.4 km s$^{-1}$ are given in the bottom panels.

\begin{figure*}
\begin{center}
\includegraphics[clip=,width= .99\textwidth]{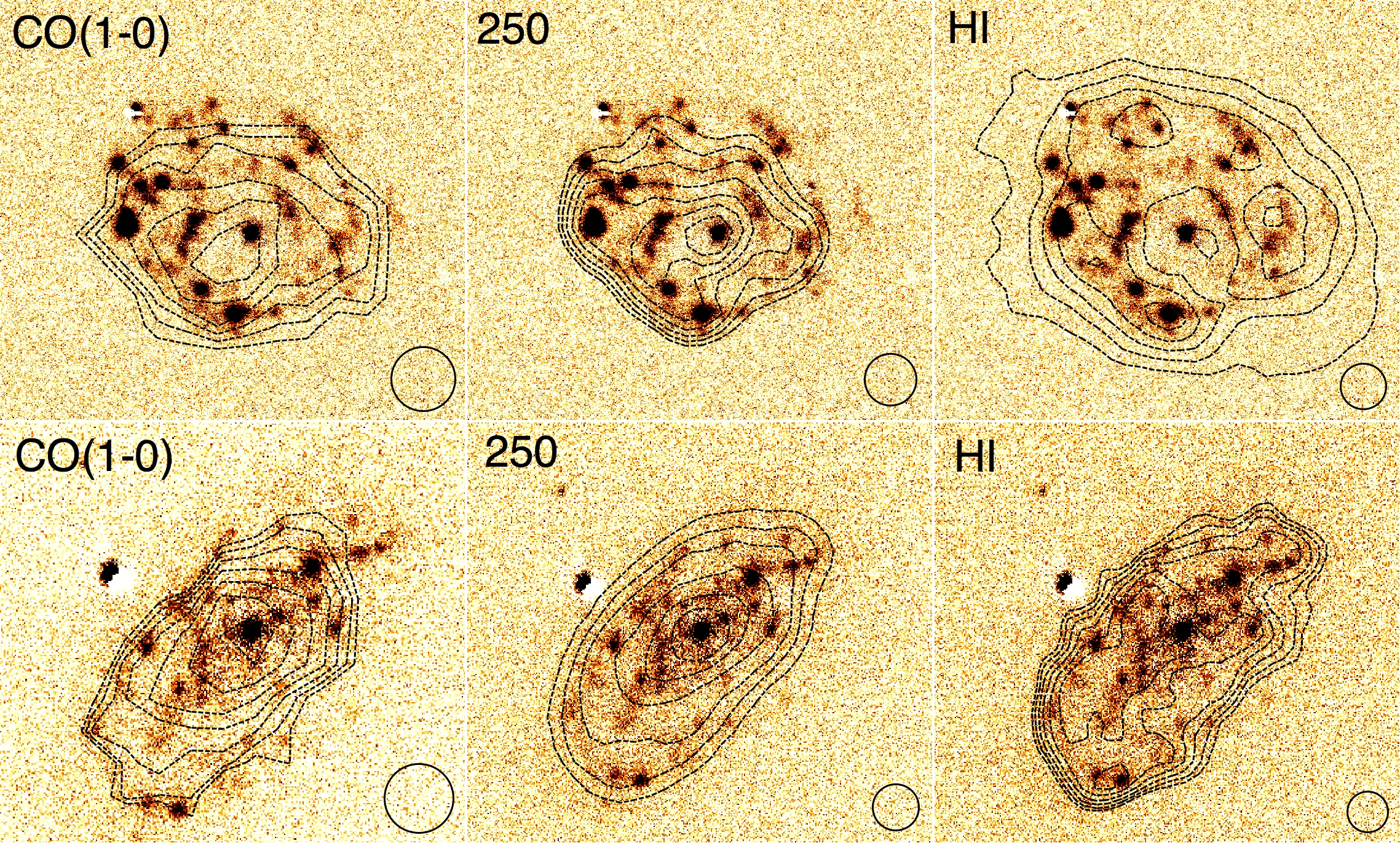}
\end{center}
\caption{H$\alpha$ emission (from GOLDMine) of NGC 4189 (top row) and NGC 4298 (bottom row) with overplotted $^{12}$CO(1-0) (left column), 250 $\mu$m emission (middle column), and HI surface mass density (right column). The contours of \couno\ line are 4.2, 6.9, 9.6, 12.3, 15, and 17.7 M$_{\odot}$ pc$^{-2}$ for NGC 4189 and 0.5, 8.8, 17.1, 25.4, 33.7, 42, and 50.3 M$_{\odot}$ pc$^{-2}$ for NGC 4298. The 250 $\mu$m contours are at 0.23, 0.28, 0.31, 0.36, 0.42, 0.48, 0.56, and 0.65 Jy beam$^{-1}$ for NGC 4189 and 0.2, 0.3, 0.41, 0.63, 0.84, and 1.06 Jy beam$^{-1}$ for NGC 4298. The HI surface mass density contours are at 4.8, 7.2, 10, 12.4, 14.8, and 17.2 M$_\odot$ pc$^{-2}$ for NGC 4189, and at 4.3, 5.2, 6, 7, 8, 9.1, 10.6 M$_\odot$ pc$^{-2}$ for NGC 4298. The beam size of each observations is shown in the bottom left of each panel.}
\label{n4189halfa}\end{figure*}

\subsection*{NGC 4189}

The CO maps of NGC 4189 are shown in Fig.~\ref{map4189}. The 3$\sigma$ noise in \couno\ and \codue\ maps is 4.2 and 4 \msolpc, respectively. The total mass of molecular gas is $M_{H_2} = 17.4\pm 1.3 \ \cdot 10^8$ M$_\odot$, and the molecular-to-total gas mass fraction is $f_{mol}$ = 0.43. The \couno\ emission is less extended ($\sim 1$\farcm6 at 4.2 \msolpc) than the atomic hydrogen ($\sim$ 2\farcm6 at 4.8 \msolpc) but comparable to the extent of the optical disk (2\farcm26). The total intensity of the $^{12}$CO(2-1) lines is comparable to that of the $^{12}$CO(1-0) line.

The $^{12}$CO(1-0) emission (top left panel of Fig.~\ref{map4189}) has a regular structure centered in the South-East region and is more homogeneously distributed with respect to the \codue\ emission, which has a clumpier structure dominated by a few unresolved regions (of 1 pixel size). Taking into account only regions with S/N $\ge$ 3 the $^{12}$CO(2-1)/$^{12}$CO(1-0) ratio varies between 0.4 and 1 (bottom right panel of Fig.~\ref{map4189}). These values are consistent with the bulk of the emission being optically thick.

In the top row of Fig.~\ref{n4189halfa} we compare qualitatively the H$\alpha$ emission of NGC 4189 \citep[from GOLDMine;][]{gav2} with the \couno\ (left), HI (middle), and the 250 $\mu$m (right) emission. Dust emission at 250 $\mu$m is comparable in size with the \couno\ emission. The distribution of dust emission across the disk is regular and more similar to the CO brightness distribution than to the HI. There is a higher concentration of the dust in the South-East regions, in agreement with a similar feature seen in the HI, \couno, and H$\alpha$ maps.

\subsection*{NGC 4298}
\begin{figure*}
\begin{center}
\includegraphics[clip=,width= .4\textwidth]{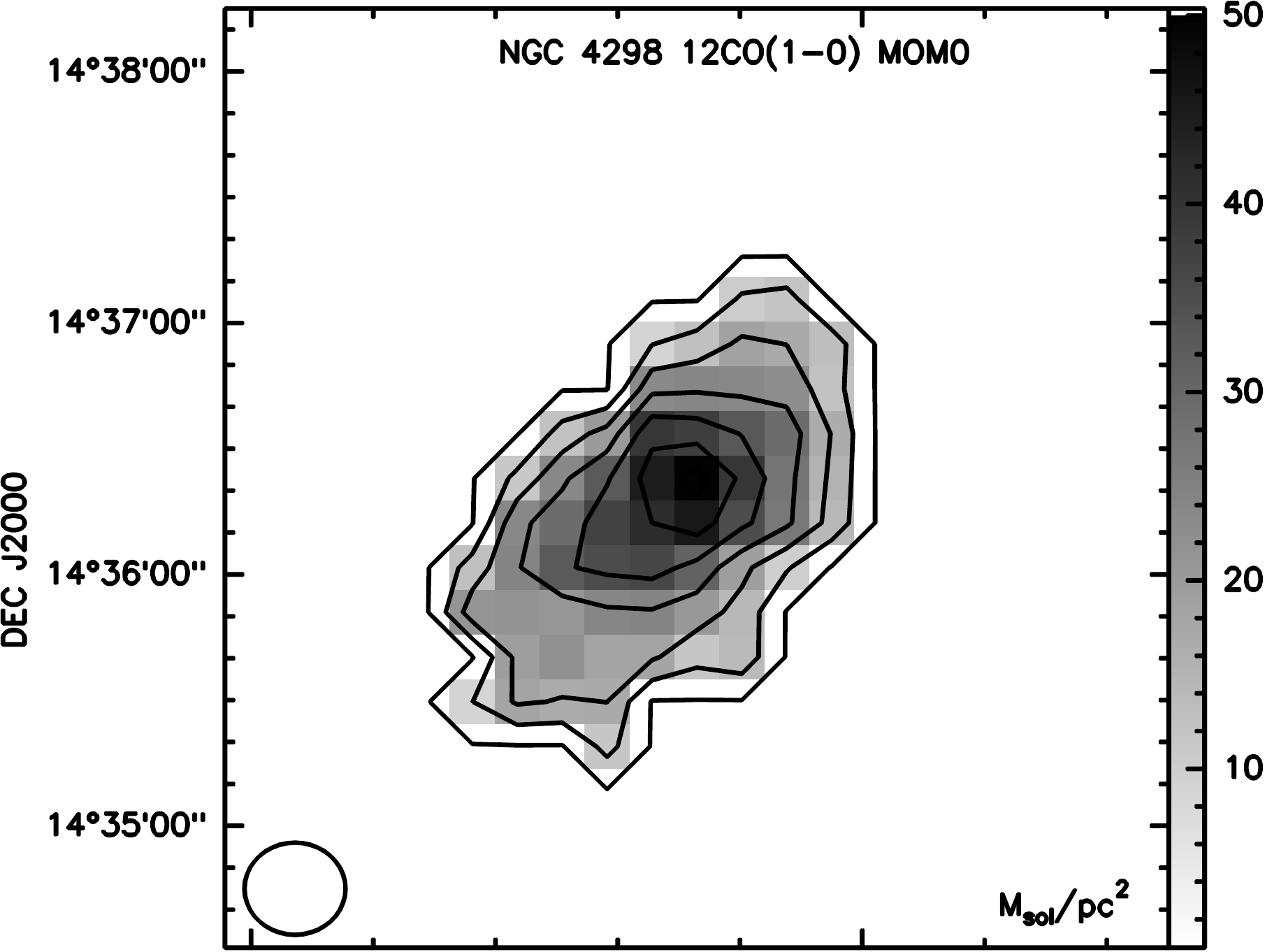}
\includegraphics[clip=,width= .343\textwidth]{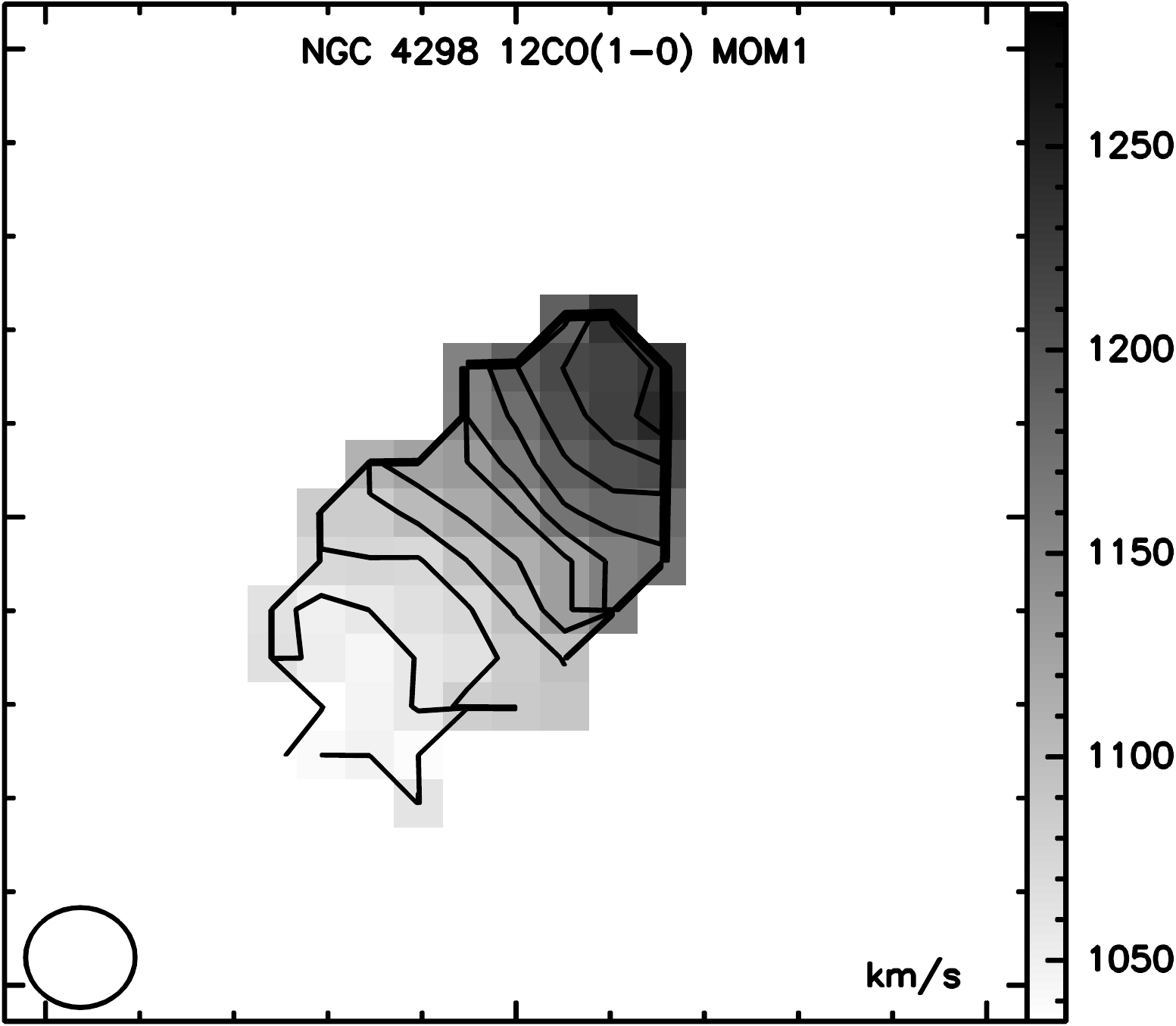}
\includegraphics[clip=,width= .4\textwidth]{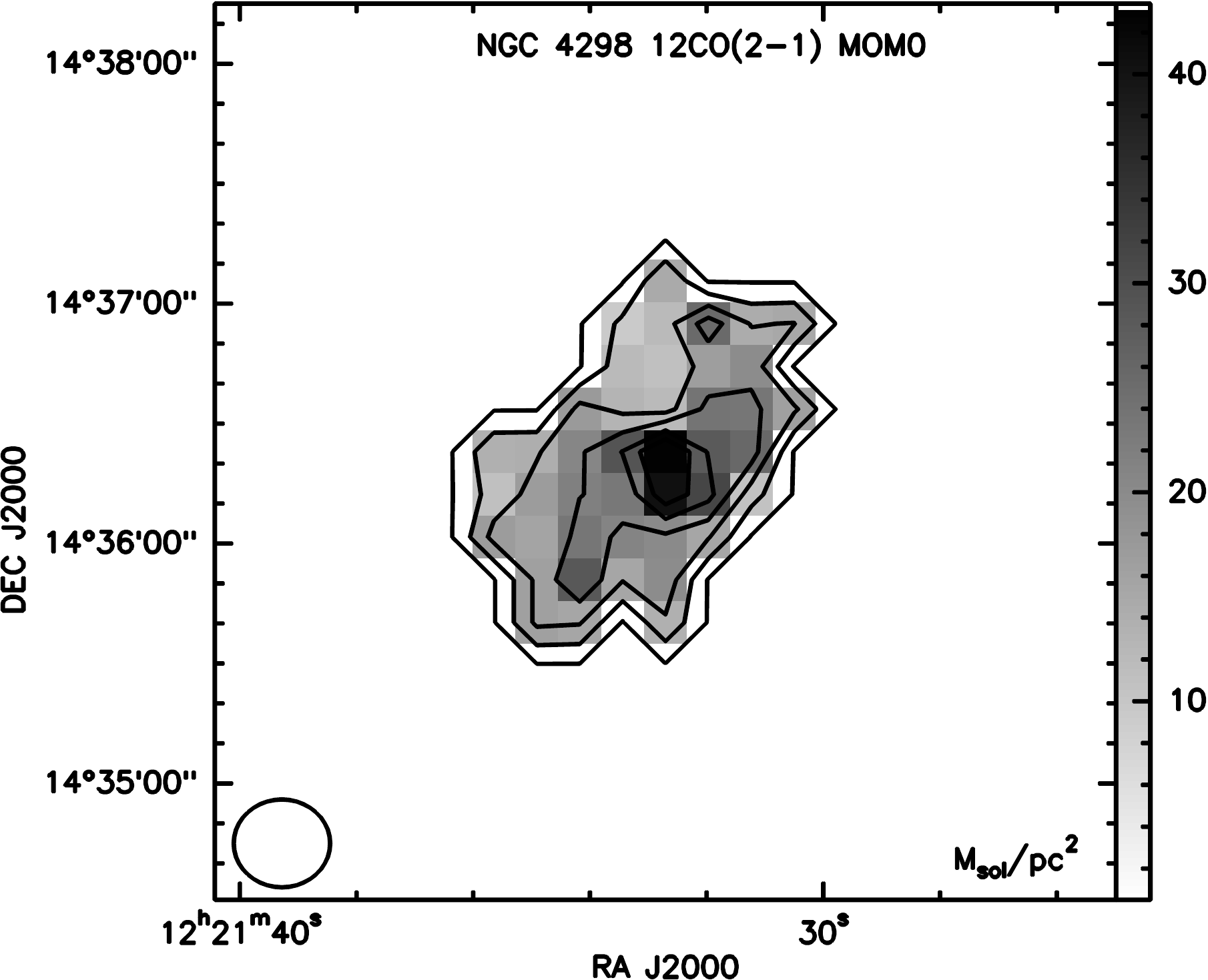}
\includegraphics[clip=,width= .35\textwidth]{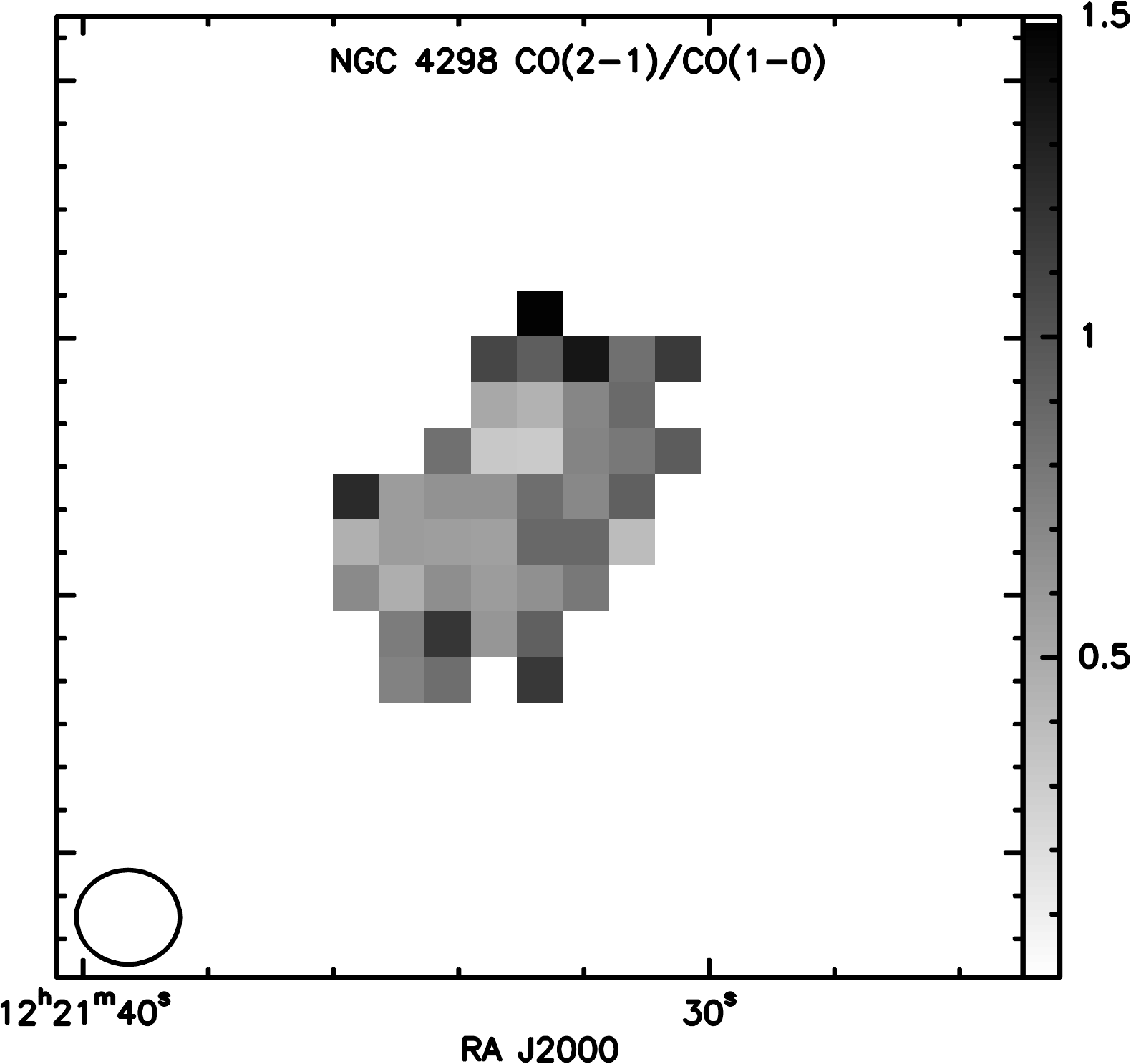}
\includegraphics[clip=,width= .69\textwidth]{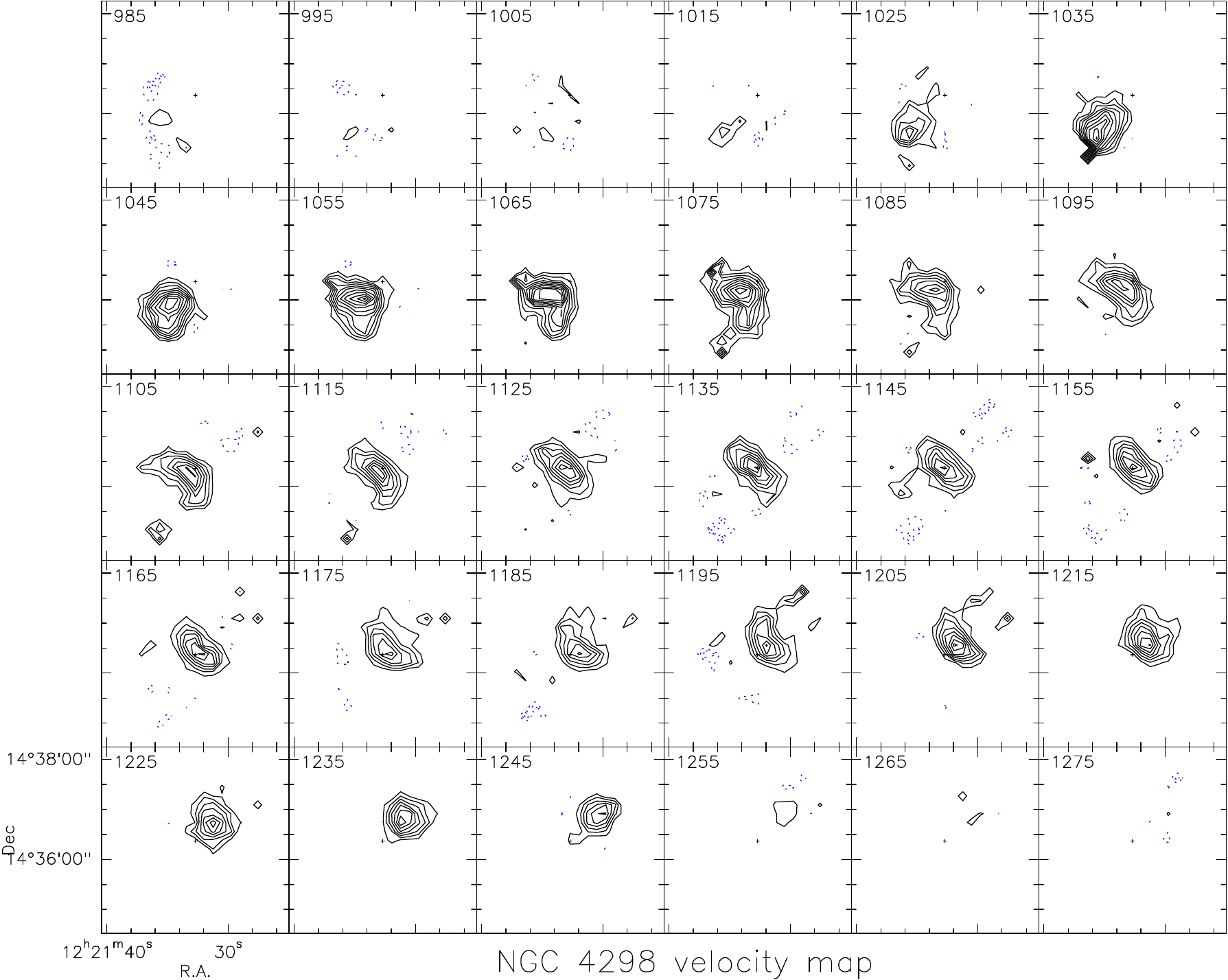}\end{center}
\caption{NGC 4298 maps. Top left: $^{12}$CO(1-0) Moment 0 map, contours: 0.5, 8.8, 17.1, 25.4, 33.7, 42, and 50.3 M$_{\odot}$ pc$^{-2}$. Top right: $^{12}$CO(1-0) Moment 1 with 1035 $\le$ v $\le$ 1235 and $\Delta$v = 20 km s$^{-1}$. Middle left: $^{12}$CO(2-1) Moment 0 map, contours: 0.5, 7.6, 14.8, 21.9, 29, and 36.2 M$_{\odot}$ pc$^{-2}$. Middle right: $^{12}$CO(2-1) over $^{12}$CO(1-0) ratio. The beam size is shown in the bottom left of each panel. Bottom panel: channels map of \couno \ with 985 $\le$ v $\le$ 1275 km s$^{-1}$ and $\Delta$v = 10 km s$^{-1}$.}
\label{map4298}
\end{figure*}

The CO maps for NGC\,4298 are shown in Fig.~\ref{map4298}. We estimate a total mass $M_{\it H_2}$ = 10.4$\pm 0.8 \cdot 10^8$ M$_\odot$ and $f_{mol}$ = 0.65. Both the $^{12}$CO(1-0) (top left panel of Fig.~\ref{map4298}) and $^{12}$CO(2-1) emissions (middle left panel of Fig.~\ref{map4298}) peak in the central regions, consistently with the observed H$\alpha$ emission (bottom row of Fig.~\ref{n4189halfa}, from GOLDMine).

The Moment 1 map (top right panel of Fig.~\ref{map4298}) shows a quite regular structure in the northern part of the disk and slightly disturbed velocity contours in the South. The $^{12}$CO(2-1)/$^{12}$CO(1-0) ratio has values between 0.3 and 1.4, with the highest values found in isolated clumps of 1-2 pixels in size. The location of these pixels corresponds to the brightest HII regions in the galaxy.

Both the HI and the CO emission have a higher surface mass density in the South-East region, but the 250 $\mu$m emission is only slightly asymmetric (see bottom row panels of Fig.~\ref{n4189halfa}).

\subsection*{NGC 4388}

\begin{figure*}[p]
\begin{center}
\includegraphics[clip=,width= .49\textwidth]{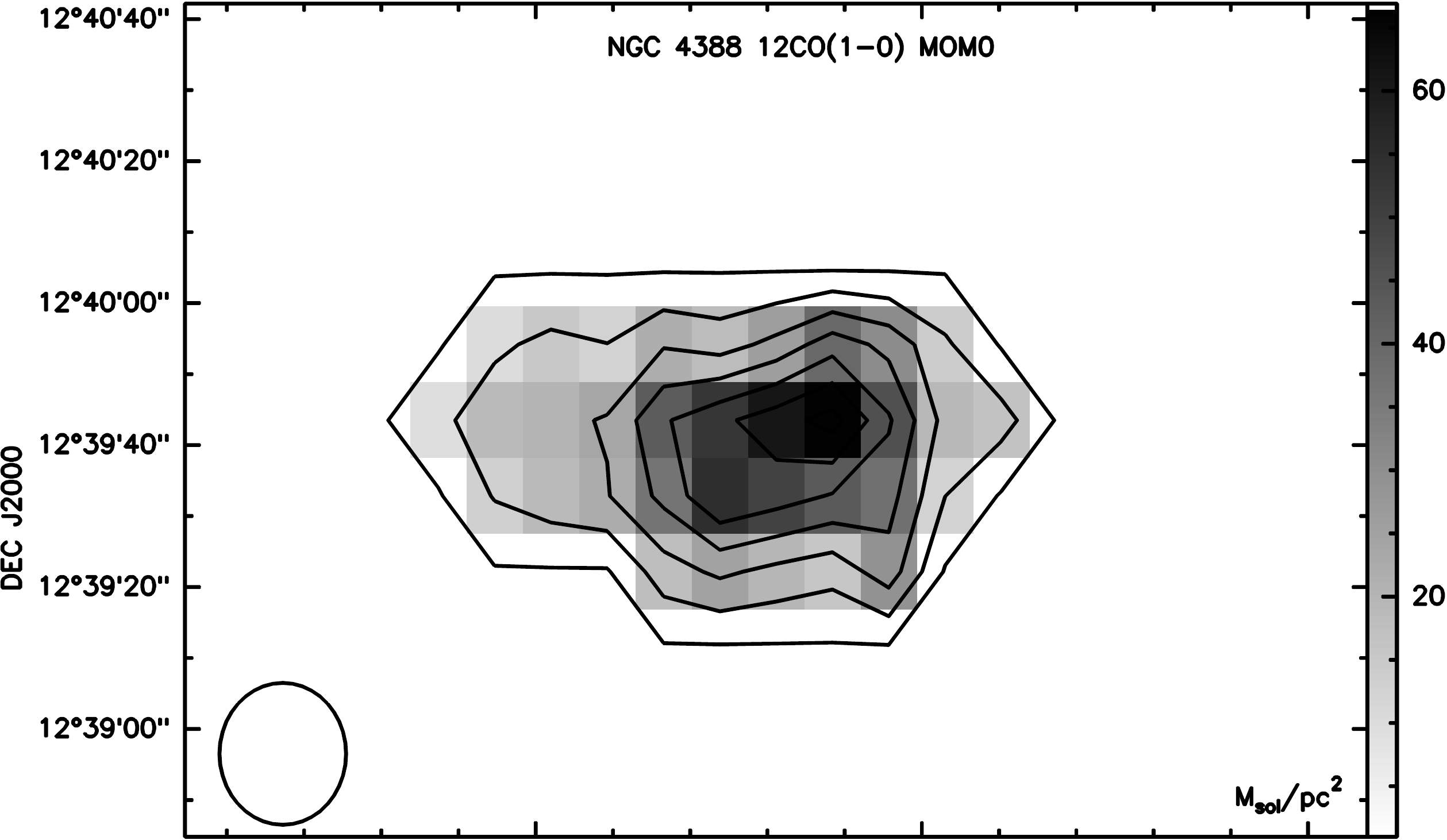}
\includegraphics[clip=,width= .44\textwidth]{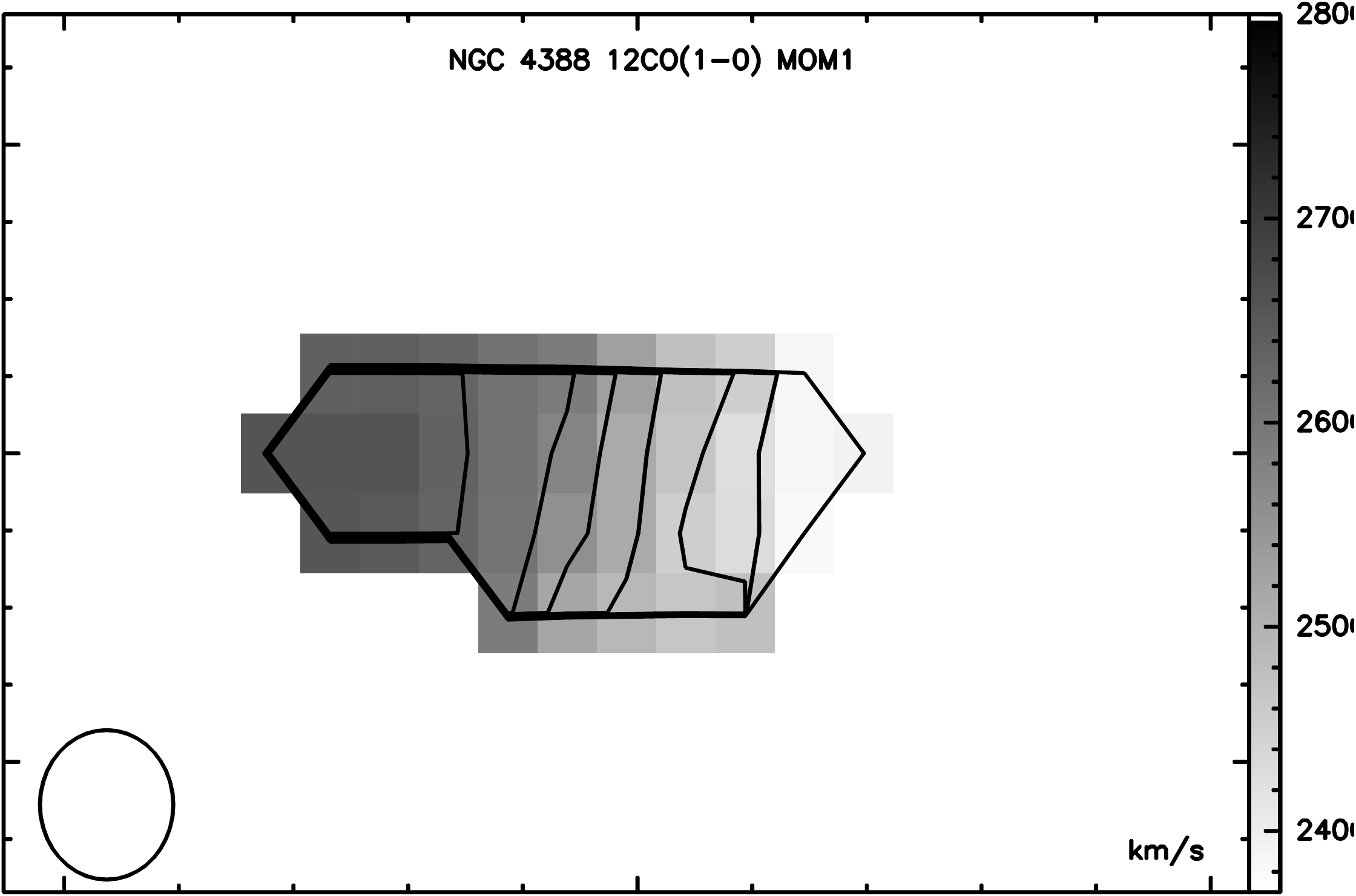}
\includegraphics[clip=,width= .49\textwidth]{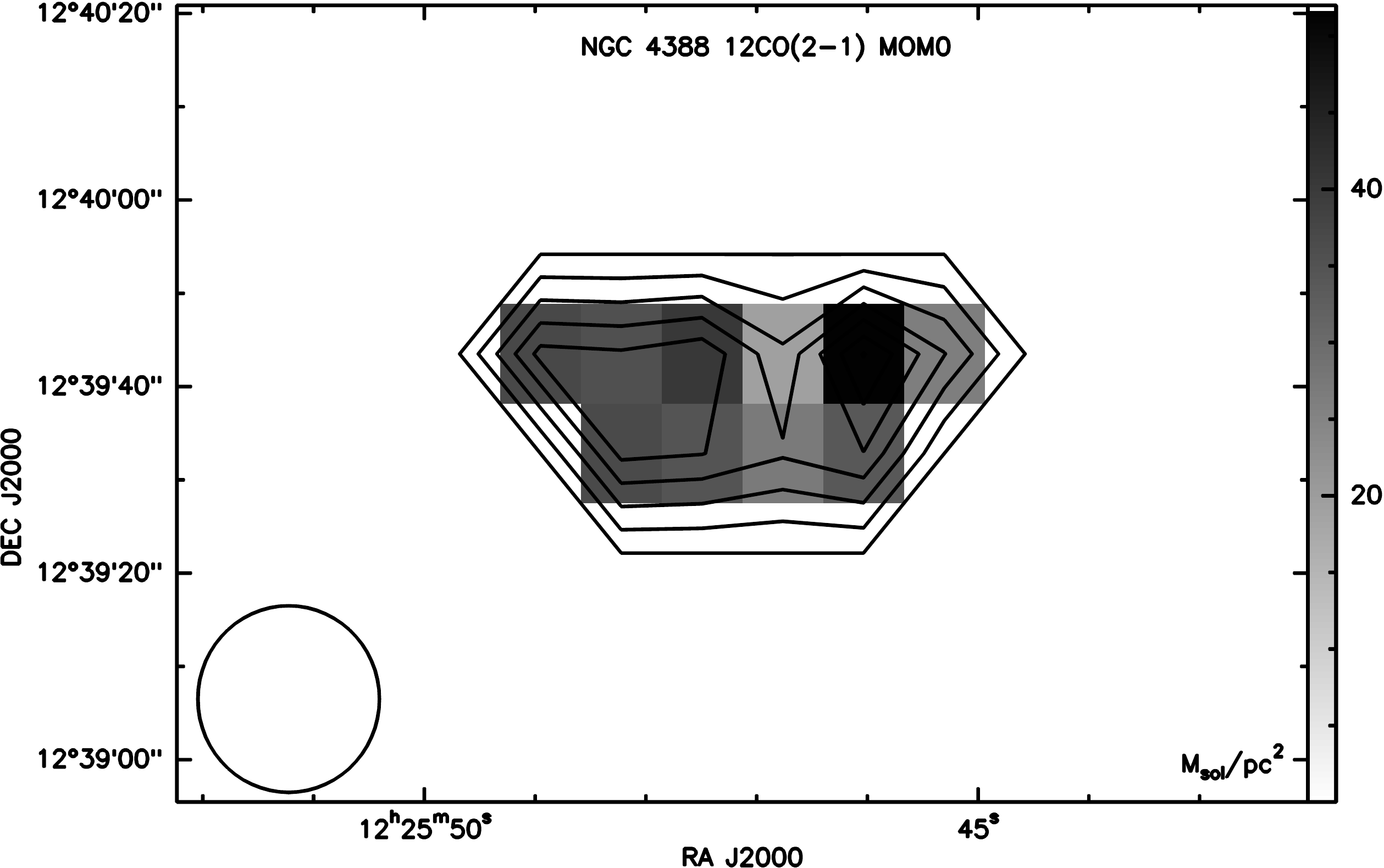}
\includegraphics[clip=,width= .44\textwidth]{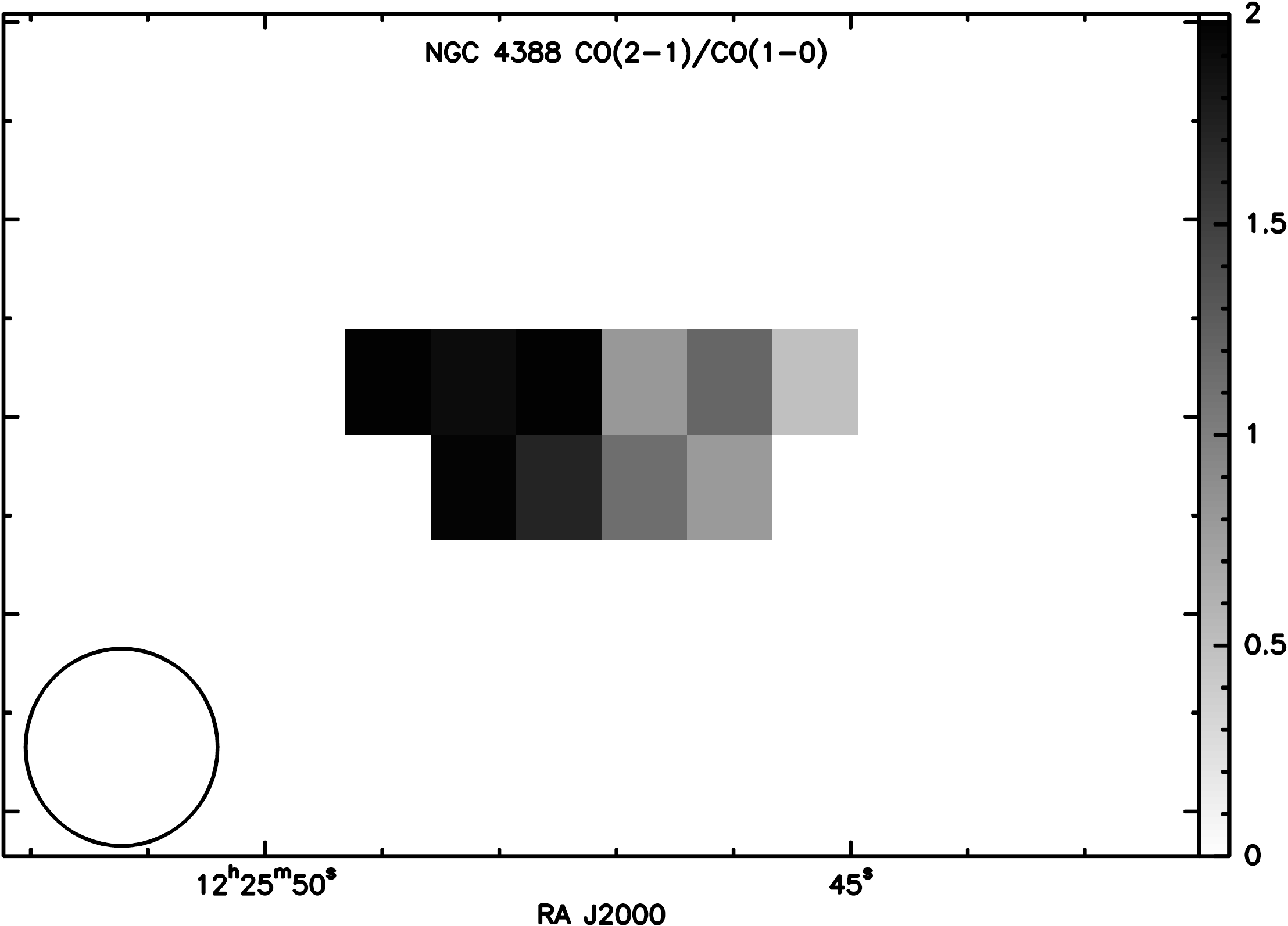}
\includegraphics[clip=,width= .69\textwidth]{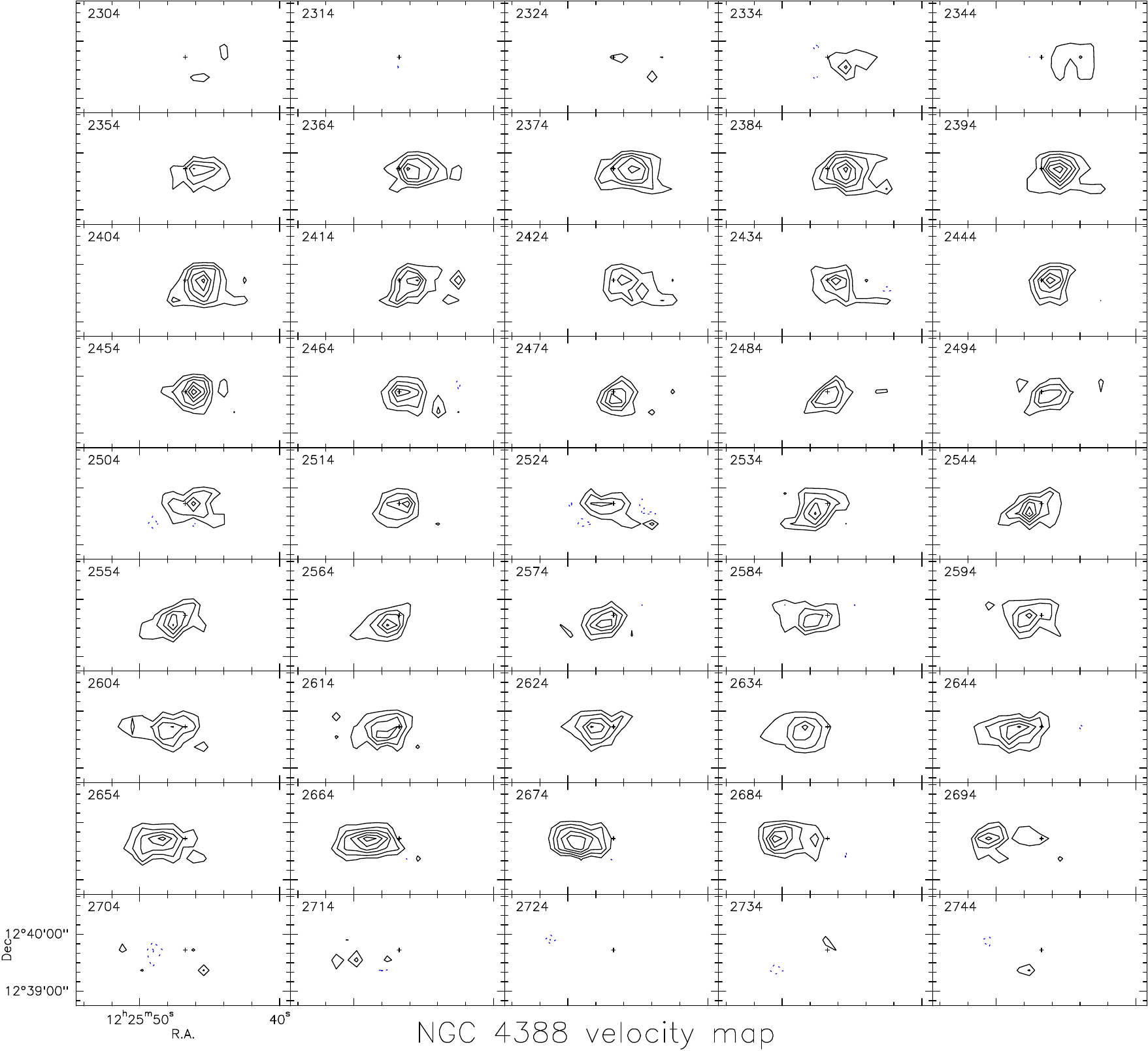}
\end{center}
\caption{NGC 4388 maps. Top left panel: $^{12}$CO(1-0) Moment 0 map, contours: 1, 12, 23, 34, 45, 56, and 67 M$_{\odot}$ pc$^{-2}$. Top right panel: $^{12}$CO(1-0) Moment 1 with 2370 $\le$ v $\le$ 2800 and $\Delta$v = 43 km s$^{-1}$. Middle left panel: $^{12}$CO(2-1) Moment 0 map, contours: 0.01, 8.6, 17.3, 26, 34.7, 43.3, and 52 M$_{\odot}$ pc$^{-2}$. Middle right panel: $^{12}$CO(2-1) over $^{12}$CO(1-0) ratio. The beam size is shown in the bottom left of each panel. Bottom panel: channels map of \couno \ with 2304 $\le$ v $\le$ 2744 km s$^{-1}$ and $\Delta$v = 10 km s$^{-1}$.}
\label{map4388}
\end{figure*}

\begin{figure}
\begin{center}
\includegraphics[clip=,width= .49\textwidth]{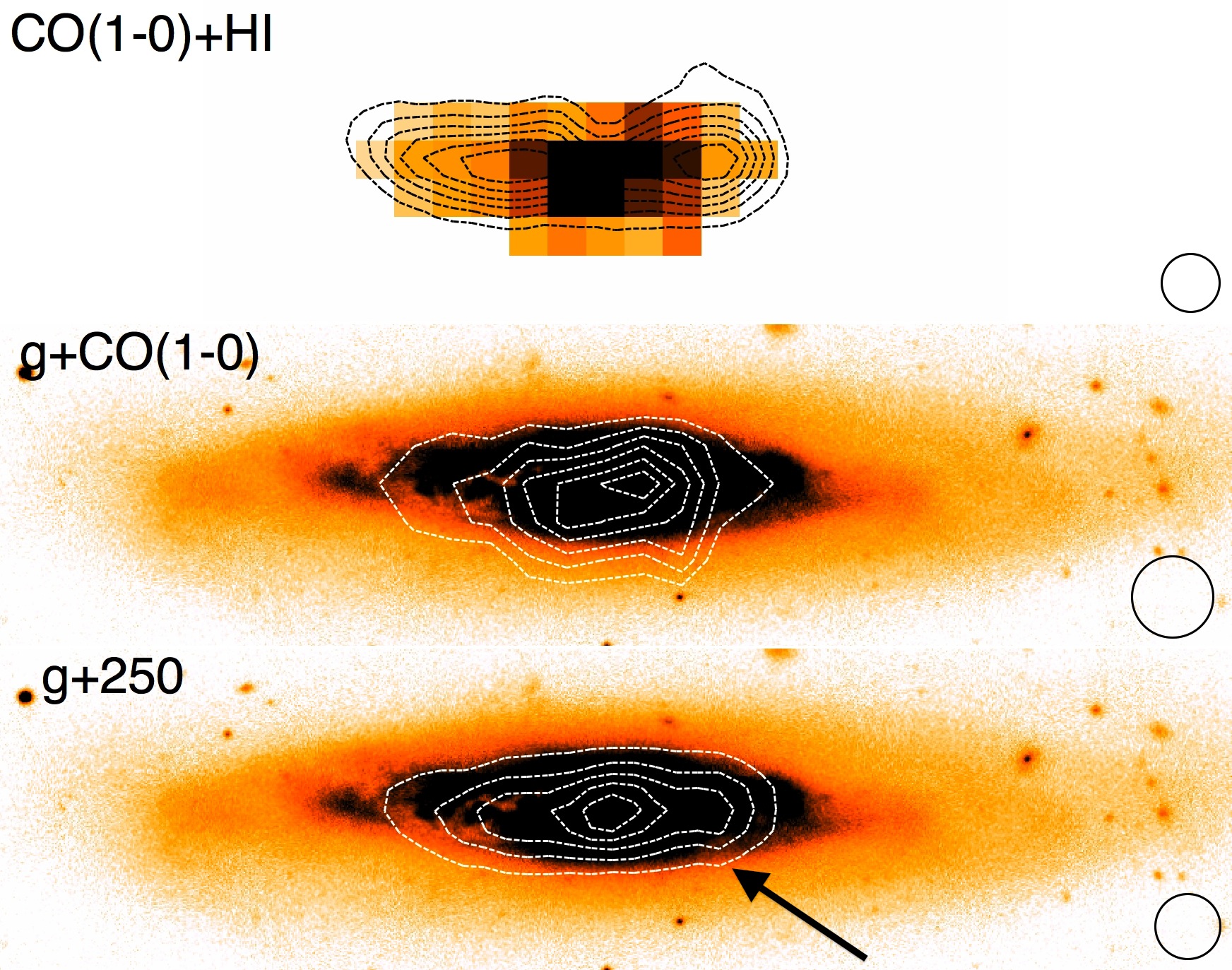}
\end{center}
\caption{Top panel: $^{12}$CO(1-0) emission of NGC 4388 with overplotted HI emission from VIVA \citep{chu}. Contours: 15.7, 18.7, 22.1, 25.2, 28.6 M$_\odot$ pc$^{-2}$. Middle panel: SDSS g-band image of NGC 4388 with overplotted $^{12}$CO(1-0) emission, contours: 1, 12, 23, 34, 45, 56, and 67 M$_{\odot}$ pc$^{-2}$. Bottom panel: SDSS g-band image of NGC 4388 with overplotted 250 $\mu $m dust emission, contours: 0.35, 0.71, 1.07, 1.43, 1.79, and 2.15 Jy beam$^{-1}$. The beam size is shown in each panel. The arrow in the bottom panel indicates the direction of ram pressure.}
\label{n4388test}\end{figure}	
\begin{figure}
\includegraphics[clip=,height=4.2cm]{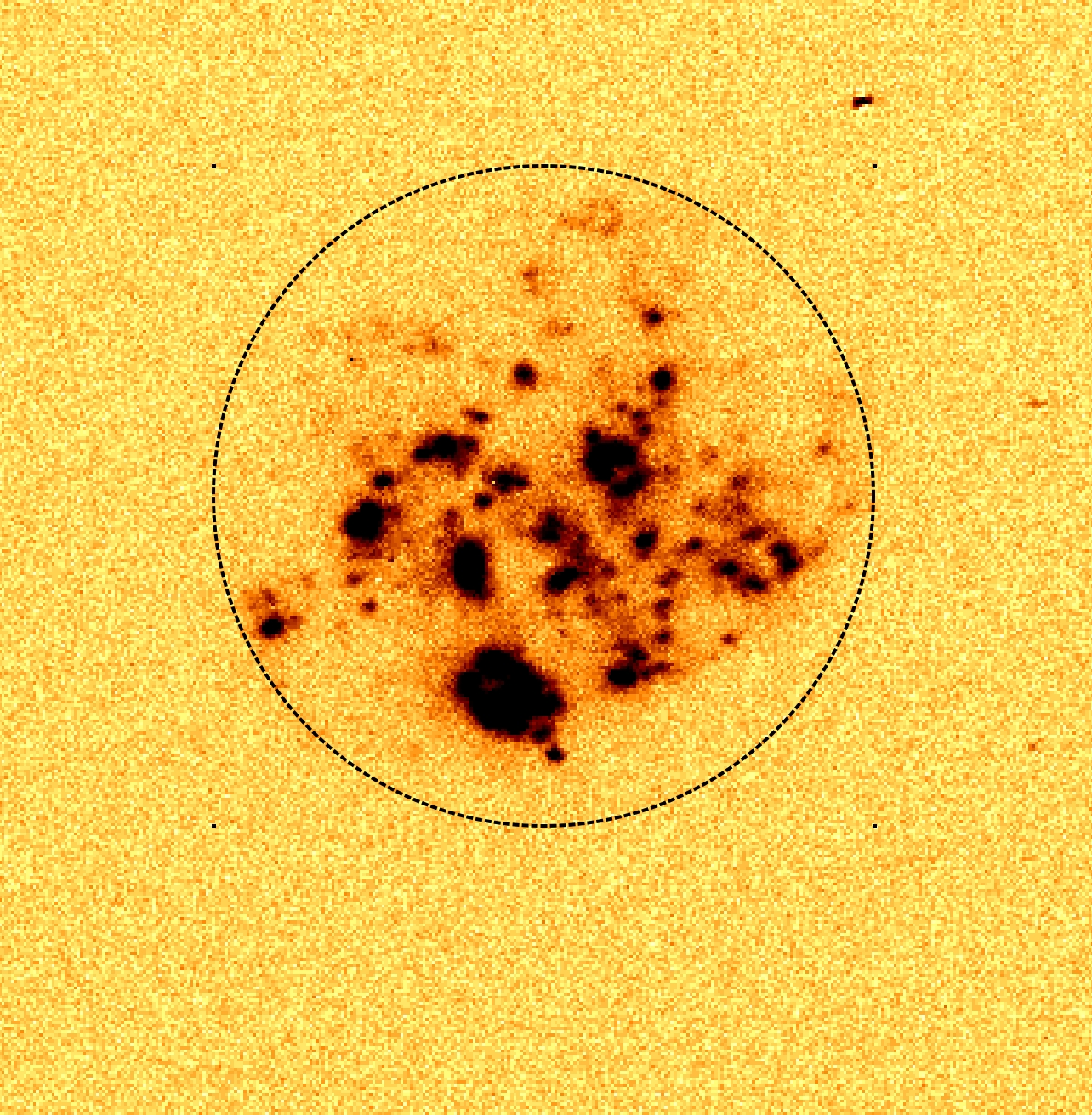}
\includegraphics[clip=,height=4.2cm]{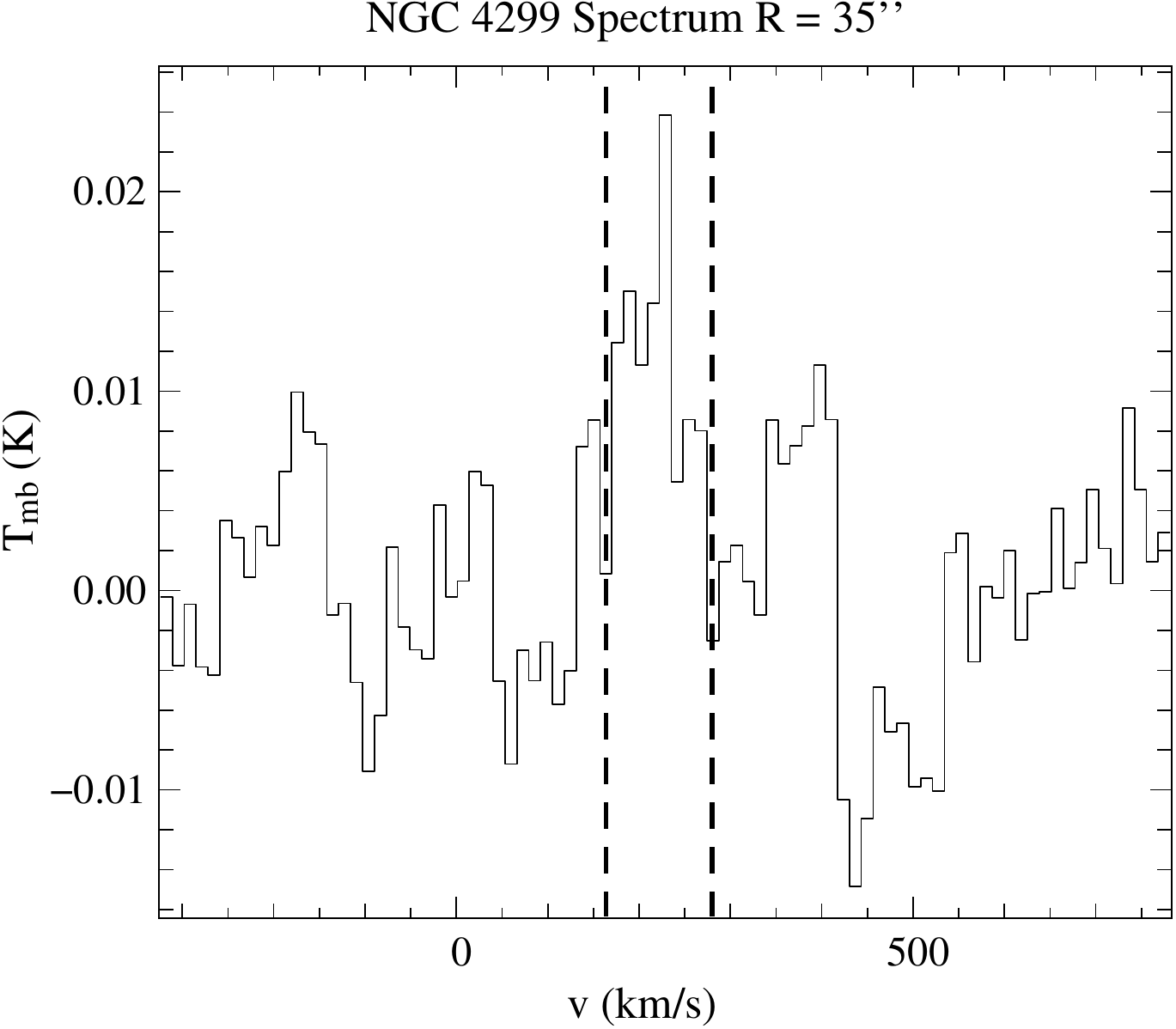}
\caption{Letf panel: H$\alpha$ emission of NGC 4299 with overplotted a circle with R = 35$''$ selected to integrate the \couno\ emission. Right panel: averaged spectra of NGC 4299 inside the circle defined in left panel. The vertical shaded lines show the velocity limits used for the integration.}\label{4299al}\end{figure}

\begin{figure}
\begin{center}
\includegraphics[clip=,height=5cm]{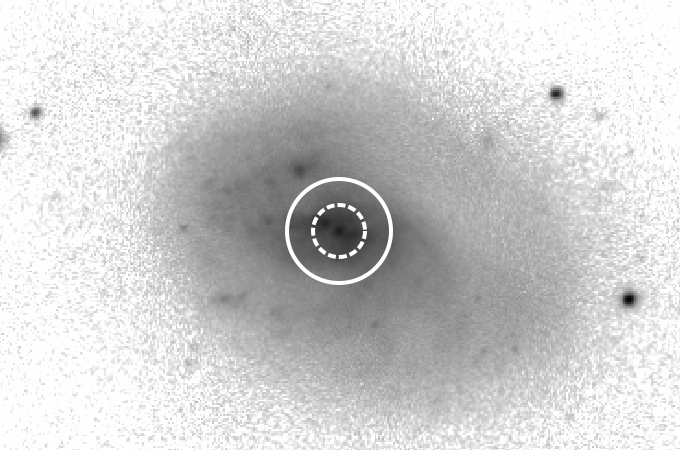}

\includegraphics[clip=,height=5cm]{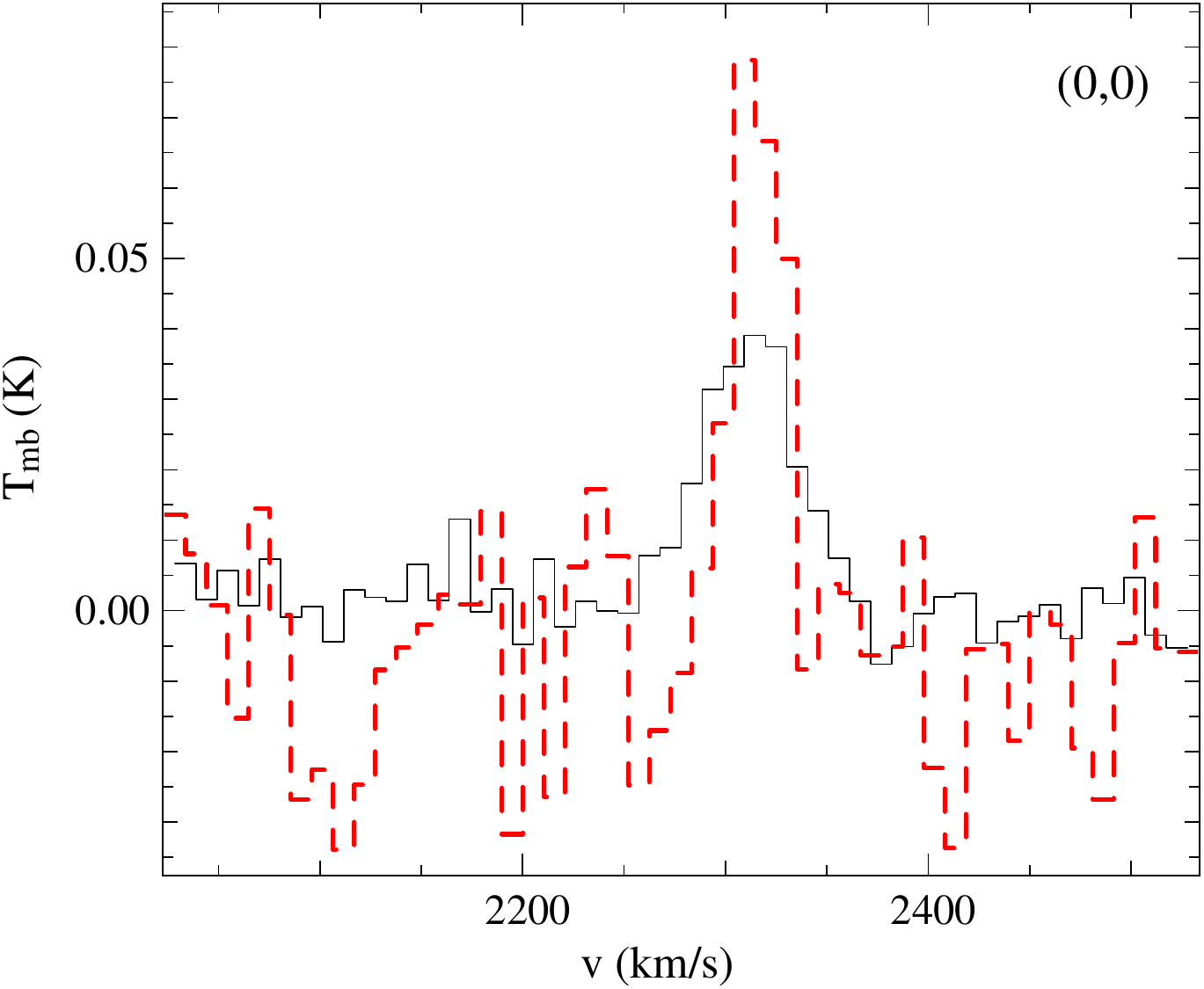}
\end{center}
\caption{Top panel: SDSS g-band image of NGC\,4351. White circles show the beam size for $^{12}$CO(1-0) (solid line) and $^{12}$CO(2-1) (dashed line). Bottom panel: $^{12}$CO(1-0) (black solid line) and $^{12}$CO(2-1) (red dashed line) observations in the position showed in the left panel.}
\label{n4351}\end{figure}

For NGC\,4388 the CO maps are shown in Fig.~\ref{map4388}. The estimated total mass of molecular hydrogen is 7.1$\pm 0.6 \cdot 10^8$ M$_\odot$ and the molecular fraction is $f_{mol}$ = 0.63. Unfortunately the $^{12}$CO(2-1) emission has SNR $>$ 3 only in the central regions and the relative map is smaller than \couno\ map. Both the HI and CO distributions throughout the disk drop sharply below the sensitivity of our observations; this can be seen in the middle panel of Fig.~\ref{n4388test}, where we overlay the contours of the molecular gas onto a reference g-band image from SDSS-DR6 \citep{ade}. The atomic and molecular components are truncated at $R/R_{25}\simeq$\,0.6, where $R_{25}$ is the radius at a $B$-band surface brightness of 25 mag\,arcsec$^{-2}$ ($R_{25} = a/2$, see column 8 of Table \ref{sample}), as expected for highly HI-deficient galaxies (\defhi \ = 1.16). The top panel of Fig.~\ref{n4388test} shows that the radial extent of the atomic and molecular phases for this galaxy are comparable. 

The structure of the $^{12}$CO(1-0) emission is asymmetric, with a more prominent elongated diffuse tail in the east side of the disk (middle panel of Fig.~\ref{n4388test}). A similar feature is present also in the HI distribution \citep{chu}, and this overdensity is higher where the gas plume observed in \citet{oos} connects with the disk. At 250 $\mu$m the disk is also slightly asymmetric, similar in shape to the CO distribution (middle and bottom panels of Fig.~\ref{n4388test}). This asymmetry in the ISM component is probably due to a recent ram pressure episode \citep[see arrow in Fig. \ref{n4388test},][]{vol,pap}. Because of the spatial limitation of our map, we were unable to investigate whether there is any stripped molecular gas associated with the elongated HI tail seen by \citet{oos}. The channel maps (top right panel of Fig.~\ref{map4388}) do not show the typical ``spider diagram" of spiral galaxies, because NGC\,4388 is highly inclined ($i = 75^\circ$) and the isovelocity contours are crowded along the line of sight.

\subsection*{NGC 4299}

Because of the low signal-to-noise ratio of the observations, the CO emission in this galaxy has been determined by integrating the spectra in the velocity range 130 $\le$ v $\le$ 300 km s$^{-1}$ as determined from the HI emission \citep{chu}, and inside a circle of 35$''$ of radius from the galaxy center (Fig. \ref{4299al}). The size of the circle was chosen to match the H$\alpha$ emission, taking into account all the star forming regions of the disk. The integration gives an intensity $I_{CO}$ = 1.28 K km s$^{-1}$, with a noise of $\sigma$ = 0.28 K km s$^{-1}$ (i.e. SNR $\sim$ 4.5). The estimated mass inside the circle is M$_{H_2}$ = 1.1 $\pm$ 0.2 $\cdot$ 10$^8$ M$_{\odot}$ with an average surface mass density of  4 M$_\odot$ pc$^{-2}$ and one of the lowest molecular mass fractions $f_{mol}$ = 0.08 (Table~\ref{dtgvalue}).

\subsection*{NGC 4294, NGC 4351 and NGC 4424}

For the galaxies observed in PS mode Table \ref{integ} reports the integrated CO line emission (col. 5, 6), the line ratio (col. 7), and the surface mass density (col. 8) for each position. The total M$_{H_2}$ mass (col. 9) was estimated assuming an exponentially decreasing CO-intensity for the gas \citep{ken}. For both NGC 4294 and NGC 4424 we estimated a scalelength of 0.1 $R_{25}$. Since NGC 4351 was observed only in the central region, we assumed a scalelength 0.2 $R_{25}$, as determined in \citet{sch} for a sample of 33 nearby spiral galaxies.

\begin{figure*}
\includegraphics[clip=,height=5cm]{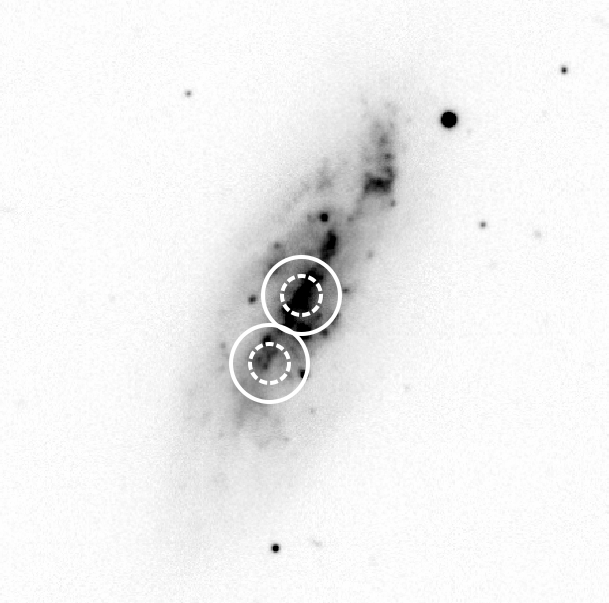}
\includegraphics[clip=,height=5cm]{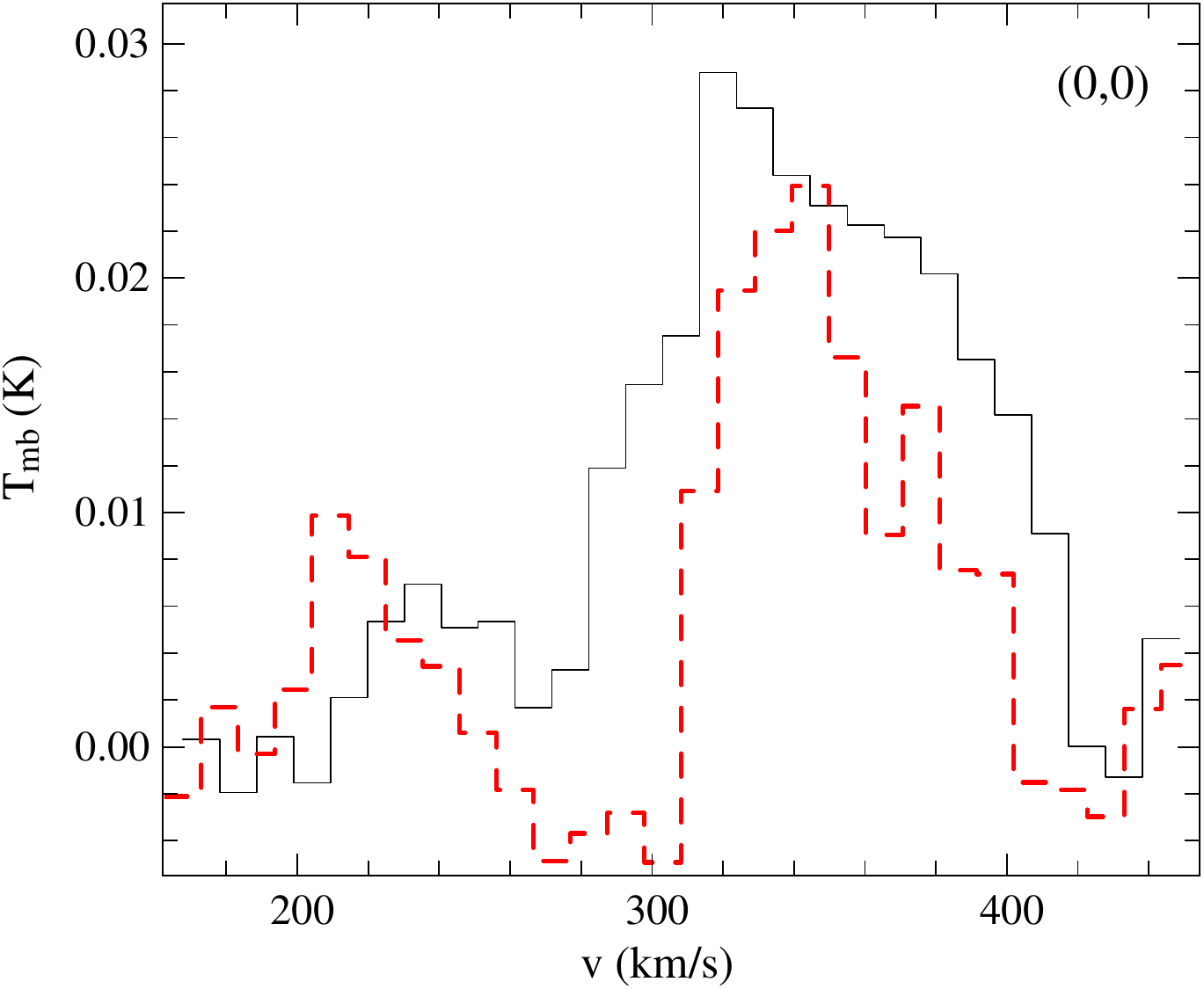}
\includegraphics[clip=,height=5cm]{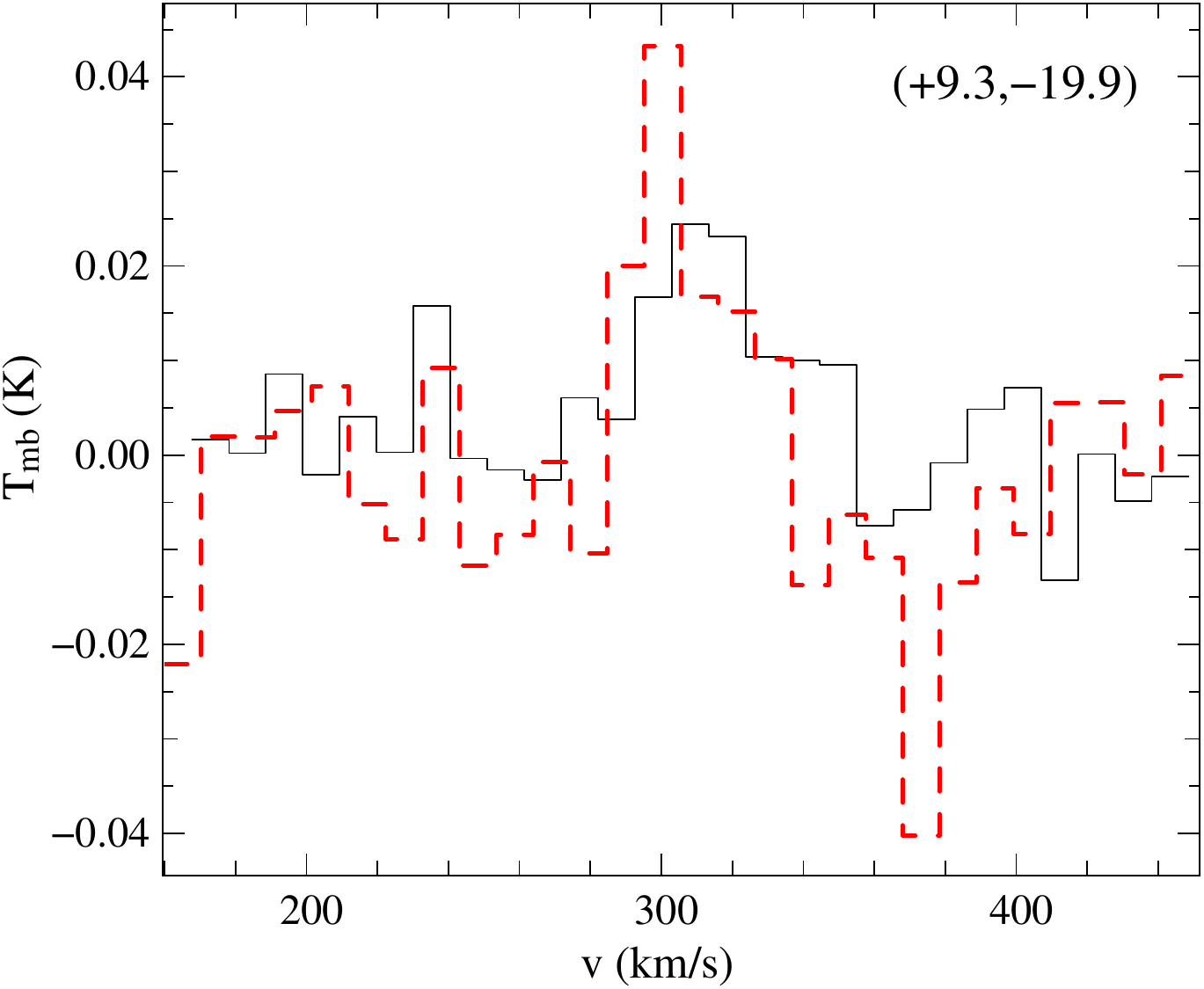}
\caption{Same as Fig. \ref{n4351} for the two positions observed in NGC\,4294. The RA and Dec offsets with respect to the center position are indicated in the top right of each spectrum panel.}\label{n4294}\end{figure*}
\begin{figure*}
\begin{center}
\includegraphics[clip=,height=4.35cm]{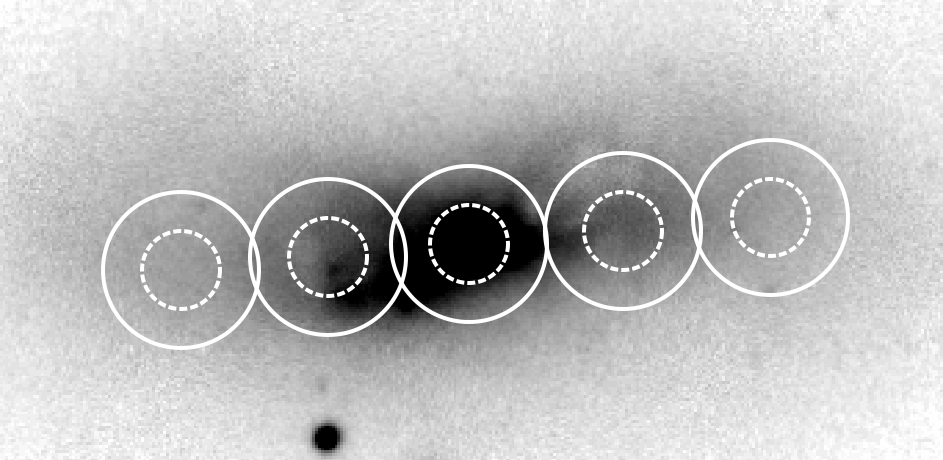}
\end{center}
\begin{center}
\includegraphics[clip=,height=5cm]{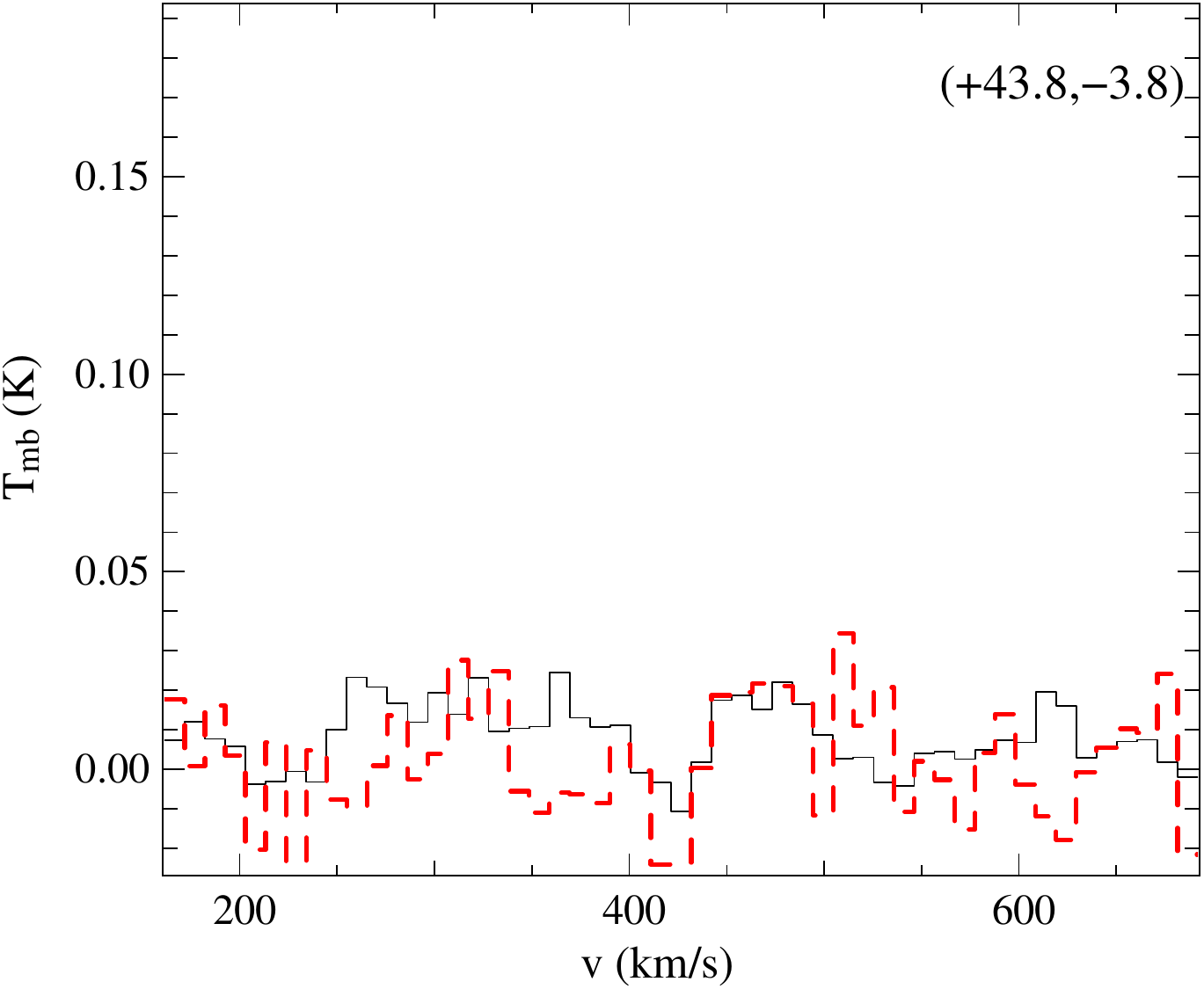}
\includegraphics[clip=,height=5cm]{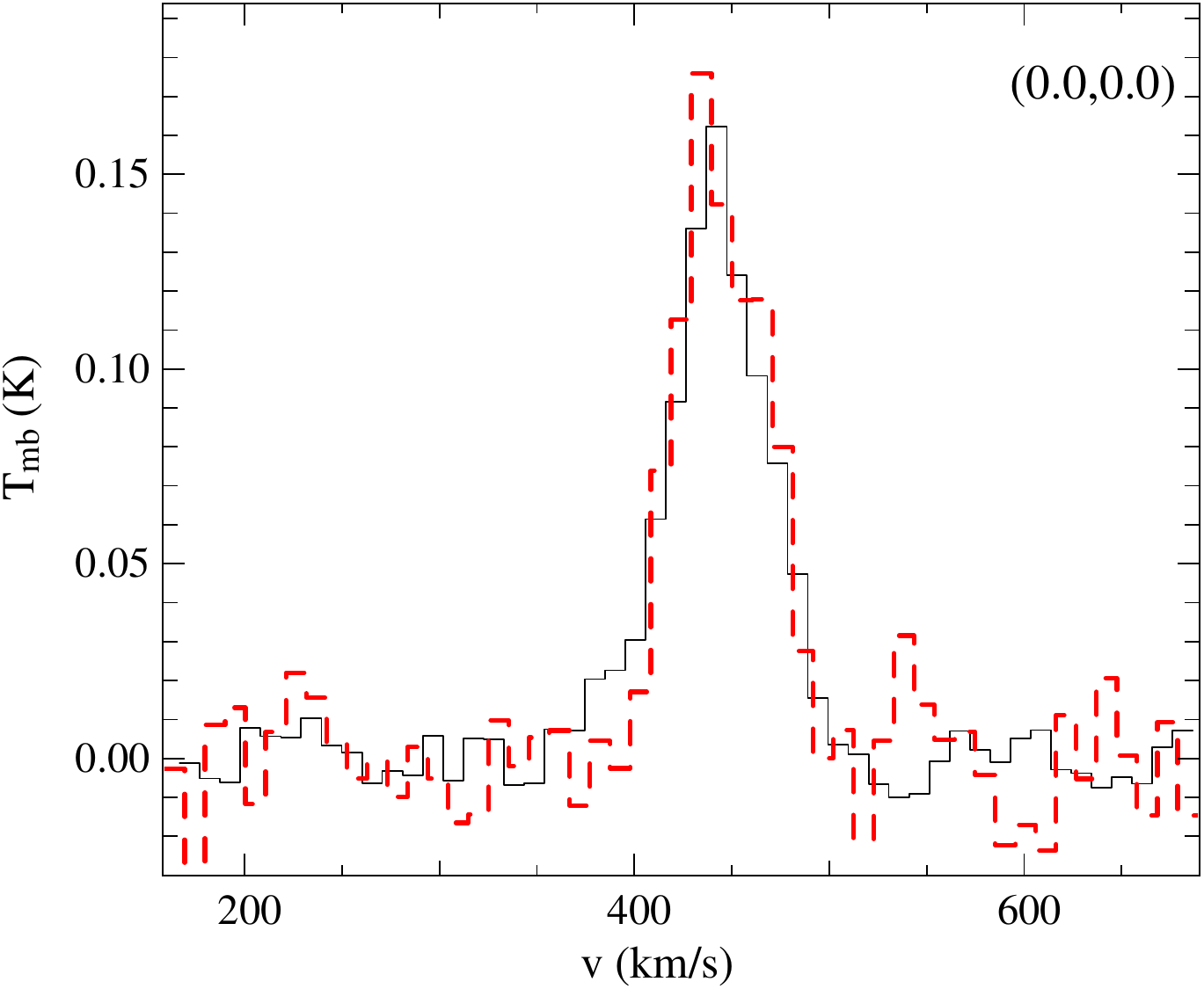}
\includegraphics[clip=,height=5cm]{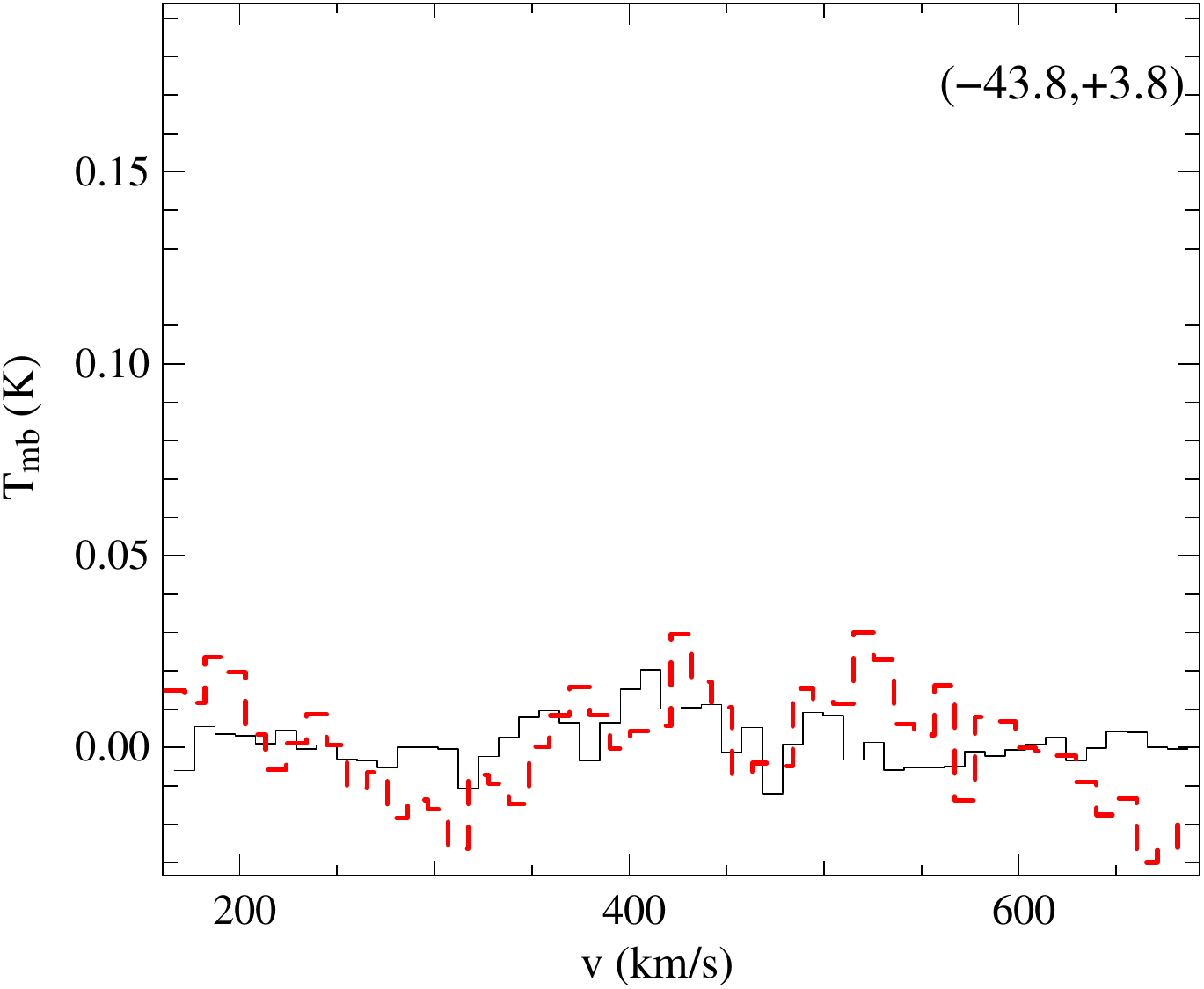}
\includegraphics[clip=,height=5cm]{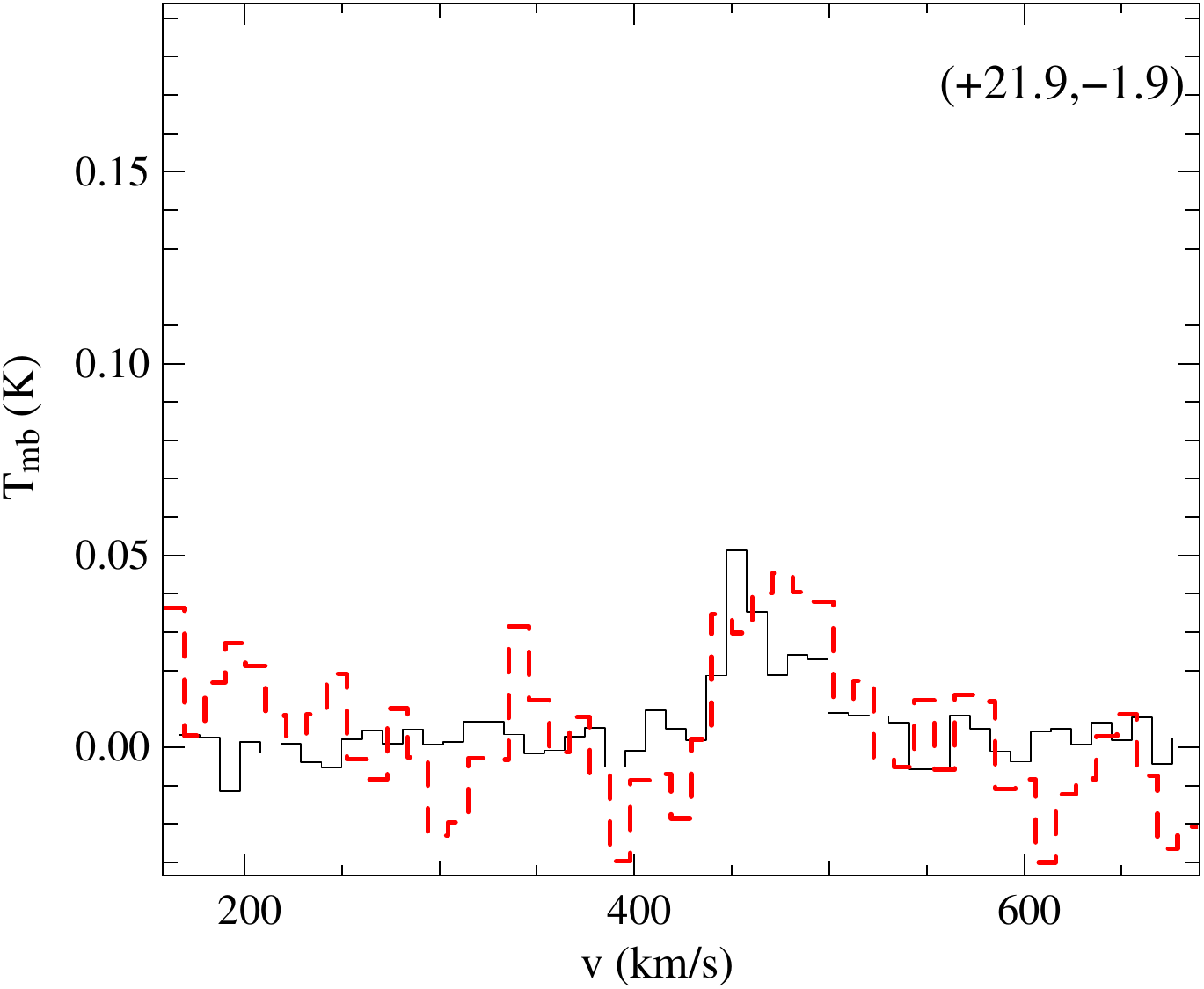}
\includegraphics[clip=,height=5cm]{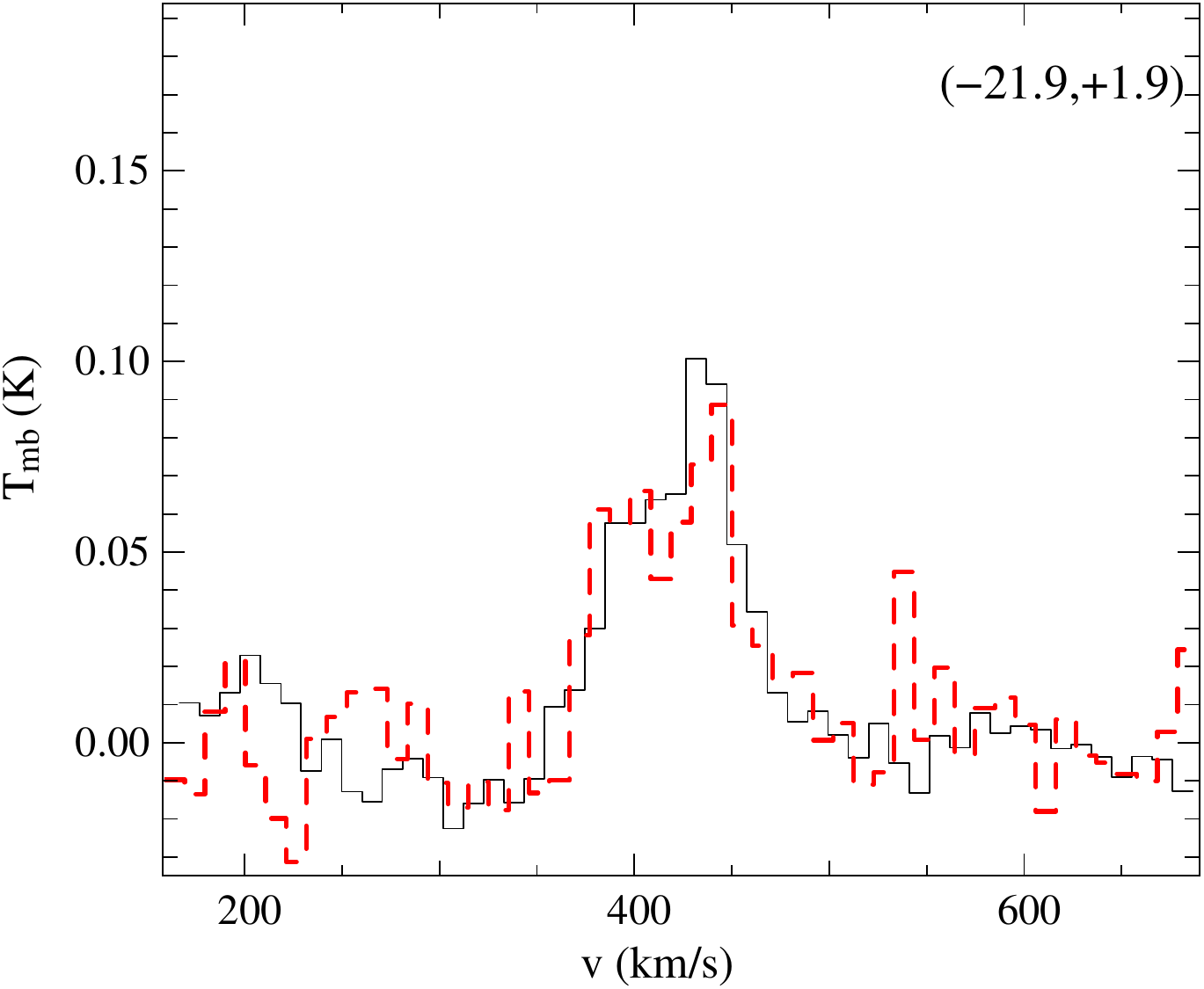}
\end{center}
\caption{Same as Fig. \ref{n4294} for the five positions observed in NGC\,4424.}
\label{n4424}\end{figure*}

Fig. \ref{n4351}, \ref{n4294}, and \ref{n4424} show the spectra with the associated positions. For NGC 4351 and NGC 4294 we found the lowest $H_2$ masses of the observed sample, 7.1$\cdot 10^7$\,M$_\odot$, and 7.3$\cdot 10^7$\,M$_\odot$, and also low molecular fractions, $f_{mol}$ = 0.17, and 0.04, respectively. For NGC 4424 we found a M$_{H_2}$ =  4.2 $\cdot10^8$ M$_\odot$, with a molecular fraction $f_{mol}$ = 0.51. The integrated fluxes of both $^{12}$CO(1-0) and $^{12}$CO(2-1) have comparable values.

\subsection{Comparison with previous molecular mass estimates}

\begin{figure}
\begin{center}
\includegraphics[clip=,width= .4\textwidth]{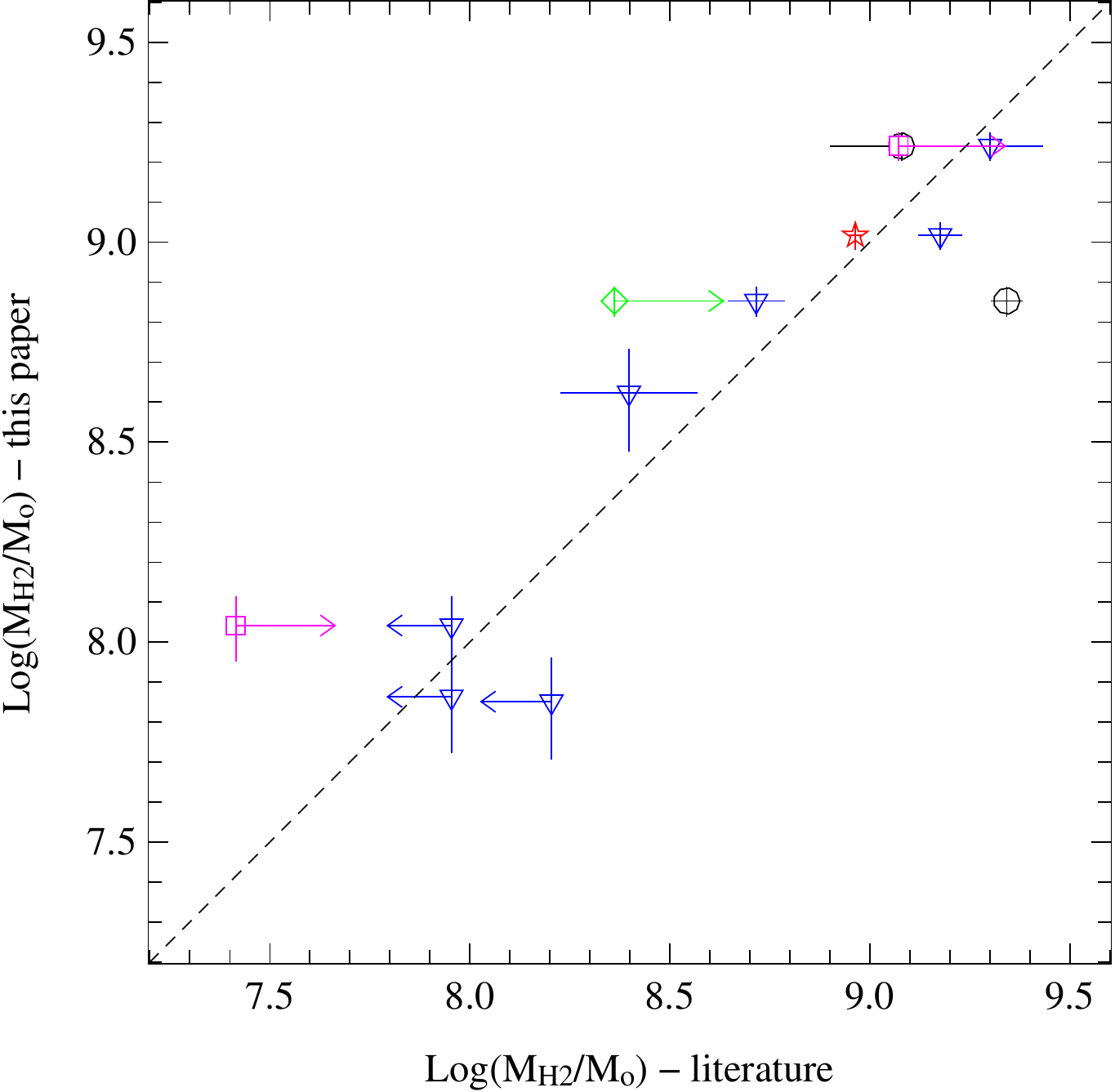}
\end{center}
\caption{Comparison of our results with previous observations. References: \citet{you} (blue triangles), \citet{chu3} (red star), \citet{sta} (black circles), \citet{smi2} (magenta squares), and 	\citet{bos6} (green diamonds). Arrows represents upper or lower limits.}
\label{compa}
\end{figure}

Figure \ref{compa} shows the comparison of our measurement of the molecular gas mass with previous estimates \citep{sta,you,bos6,smi2,chu3}. We have rescaled the literature values to match our choice of \xco \ and distance. Overall, there is a quite good agreement with previous observations. The only exception is NGC 4388 for which \citet{sta} reported M$_{H_2}$ = 2.2$\cdot10^9$ M$_\odot$, a value $\sim$30$\%$ higher than our estimate. The beam size of their telescope is larger, 100$''$, and they observed the disk in 5 different positions, with an offset of 1\farcm5 in East-West and North-South directions. Surprisingly they found in the East and West side of the disk an emission comparable to the value in the center of the disk. Our map does not extend farther than 80$''$ along the disk; however the emission decreases steeply already at $\sim $30-40$''$ from the center. Our molecular mass estimate agrees with the value of \citet{ken2} which have sampled similar regions of this galaxy as \citet{sta}, but with a smaller beam. 

\cite{cort} carried out interferometric observations of the \couno emission line of the central regions of NGC 4424, and found a molecular gas mass of 1.6$\cdot 10^8$\,M$_\odot$. Given the larger area covered by our observations, this is compatible with the mass value we find in the central beam and with our higher total mass value. 

Our observations have a higher sensitivity and better spatial resolution than previous ones. For most of the galaxies, $\sigma_{1-0}$ is between 4 and 11 mK. Moreover all the galaxies observed in OTF mapping mode have not been mapped previously.

\section{Environmental effects on the radial distribution of gas and dust}
\label{disk}

In this section we study the influence of the environment on the surface mass density distributions for the molecular, atomic, and dust components. We consider the new mapped galaxies together with 8 additional galaxies observed at the Nobeyama 45-m telescope \citep{kun}. The beam size of this set of observations is 15$''$, comparable to the IRAM-30m telescope beam (22\farcs5 for \couno \ line), but the typical rms noise level is 40-100 mK, a factor of ten higher than the level achieved by our observations.

We divide our galaxies in three groups: non-deficient galaxies with \defhi $<$ 0.4 (NGC 4299, NGC 4254, NGC 4189, NGC 4321), galaxies slightly HI-deficient, with 0.4 $\le$ \defhi \ $\le$ 0.7 (NGC 4298, NGC 4535, NGC 4192, NGC 4501), and HI-deficient galaxies, with \defhi $>$ 0.7 (NGC 4402, NGC 4579, NGC 4388, NGC 4569). We use this classification to characterize the environmental effects in our sample.

We consider the total gas surface density as the sum of the atomic and molecular gas component, since we did not take into account the helium contribution. To scale the quoted surface mass density for helium, one should multiply the recovered values by a factor $\sim$ 1.36. We also neglected the ionized gas, which is expected to be a small fraction of the total gas mass.

\subsection{Radial profiles}
\label{radialprofiles}

\begin{figure*}\begin{center}
\includegraphics[clip=,width= .9\textwidth]{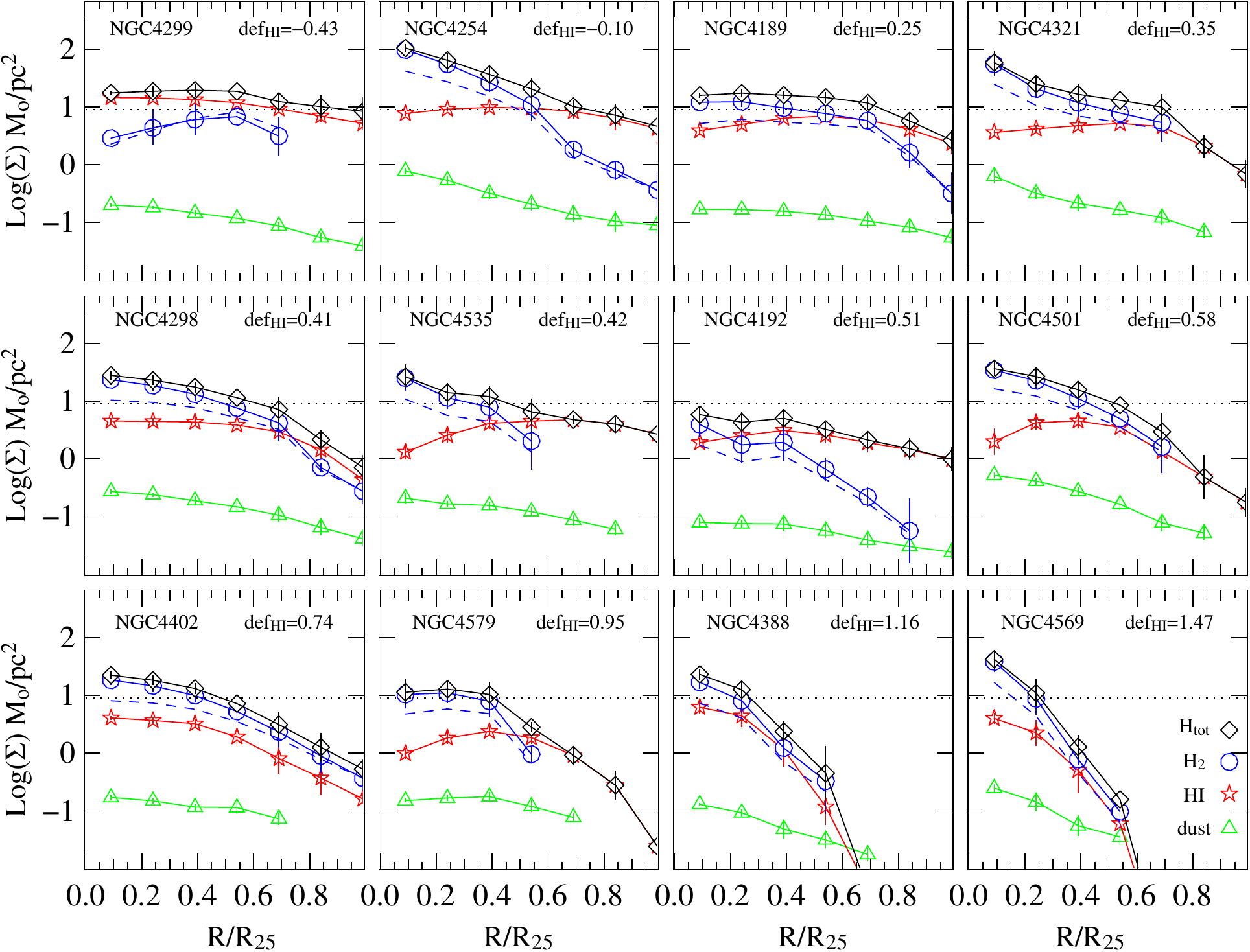}
\end{center}
\caption{Dust (green triangles), atomic gas (red stars), and molecular gas (blue circles) surface mass densities as a function of $R/R_{25}$ for the extended sample. The molecular component has been estimated using the Galactic \xco\ (solid line) and the radially varying \xco\ (Eq. \ref{scaleR}; dashed line). The total gas mass surface density (for the Galactic \xco\ only), is indicated with black diamonds. The horizontal dotted line indicates $\Sigma$ = 9 M$_{\odot}$ pc$^{-2}$.}
\label{mass2}\end{figure*}

The radial profiles are obtained averaging surface mass density in elliptical annuli of width 0.15 $R/R_{25}$, using the inclination and the major axis position angle given in columns 10, 11 of Table \ref{sample}. The average value in each annulus is plotted when the total number of pixels with surface mass density above the threshold value is at least 20\% of the total number of pixels inside the annulus. Finally the radial profiles are corrected to face-on values.

Fig. \ref{mass2} shows the radial profiles of the gas and dust components. The panels are ordered according to the \defhi \ parameter of each galaxy: the top, middle, and bottom rows refer to the non-deficient, slightly HI-deficient, and HI-deficient galaxies, respectively. The open blue circles refer to the molecular surface mass density for a constant Galactic \xco\ value. The radial profile of the atomic component is plotted using red star symbols. Since the beam size of the HI observations is similar to that of the CO observations, we kept all images at their original spatial resolution ($\sim$ 20-25$''$). The total gas profile is shown with black diamonds. The triangles connected by the green solid line show the dust surface mass density at the resolution of 36$''$.

As a general trend, the molecular gas has a higher surface mass density than the HI gas in the inner regions. In the external regions the mass surface density of H$_2$ decreases, and the relation between the total gas and the HI distribution depends on the environmental effects undergone by the galaxy. The dominance of the molecular gas in central regions is consistent, for some galaxies in our sample, with the results obtained in a sample of 33 nearby galaxies by \citet{big}. By analyzing the surface mass density of the different gas phases they recovered a threshold at which most of the gas is converted into molecular form, i.e. 9 M$_{\odot}$ pc$^{-2}$ \citep[see also][ for theoretical modelling of this effect]{kru2,kru}. In Fig.~\ref{mass2} the horizontal dotted line marks this value of surface mass density. In the central regions of NGC\,4535, NGC\,4501, and NGC\,4579 the molecular phase is already dominant where the HI surface mass density is only 1 \msolpc. However, these are HI-deficient galaxies and variations of the threshold value may occur. NGC\,4299, the galaxy with the lowest metallicity in our sample, shows a lack of CO. In the central regions the HI dominates the total gas surface mass density, even if for this galaxy it is above the threshold value.

The estimate of the molecular gas profile strongly depends on the assumption for the \xco \ factor. We have used so far the Galactic solar neighborhood value, but there is ample evidence that \xco \ varies in other galaxies as a function of physical conditions \citep[e.g.][]{mal,isr,bar,bos2,isr2,isr3,bol,nar,ler,she,nar2,bel}. Thus, we consider an alternative conversion factor, which depends on the metallicity, and varies radially following the metallicity gradient along the disk:
\begin{equation}
log\Big(\frac{X_{\rm CO}}{2 \cdot \ 10^{20}}\Big) = 0.4 \cdot \frac{R}{R_{25}}-Z_c + 8.69,
\label{scaleR}
\end{equation}
\noindent
with $Z_c$ equals to the oxygen abundance in central regions, expressed in the usual form 12+log O/H. Since we do not have $Z_c$ and gradients for all the galaxies in our sample we assume the central metallicity estimated using the mass metallicity relation \citep{tre} and reported in col. 12 of Table \ref{sample}. We fix the solar value at $Z_\odot$ = 8.69, as recently found by \citet{asp}, and assume a radial metallicity gradient $\varpropto -0.4 \ R/R_{25}$, as found by \citet{mag} for a sample of Virgo spirals. This slope should reproduce the effect of a metallicity gradient. 

The blue dashed lines in Fig. \ref{mass2} represent the surface molecular mass density calculated according to Eq. \ref{scaleR}. The choice of a radially varying \xco\ as a function of metallicity affects significantly the radial profiles and can reduce the molecular mass content up to 60\%. Considering for our sample a typical metallicity $Z_c$ = 9.09 and according to Eq. \ref{scaleR}, we obtain for the central regions \xco = 0.78 $\cdot$ 10$^{20}$ [K km s$^{-1}$]$^{-1}$, considerably lower than the Galactic \xco. For NGC 4189, if we use the galactic \xco\ we obtain a centrally peaked molecular distribution with a surface mass density 2-3 times larger than HI in the central region. For $R/R_{25} \gtrsim$ 0.7, when the surface mass density is below 4 M$_\odot$ pc$^{-2}$, the HI phase dominates over the molecular phase. Rescaling the conversion factor as in Eq. \ref{scaleR}, because of the high value of $Z_c = 9.09$, the atomic hydrogen becomes the dominant component already at $R/R_{25} \gtrsim$ 0.3, and the surface mass density of the molecular gas is never above 6 M$_\odot$ pc$^{-2}$. For NGC\,4298 the molecular gas in the central region is always dominant for every choice of the conversion factor, and the effect of a rescaling is to reduce the surface mass density value according to the metallicity. For both NGC\,4189 and NGC\,4298 the effect of a radially varying conversion factor is negligible in the external regions, $R/R_{25} >$ 0.6, where, according to our assumed gradient, the metallicity approaches solar. For NGC\,4388 the molecular component with a variable \xco \ has surface mass densities comparable to the atomic component, as a consequence of the strong HI-deficiency (\defhi = 1.16). For NGC\,4299 (top left panel of Fig. \ref{mass2}) the effect on the total gas is negligible, because the galaxy is HI dominated (\defhi = -0.43).

The radial profiles of dust surface mass densities are shown as green triangles in Fig. \ref{mass2}. The slope of the dust profiles tends to be intermediate between the molecular component (steeper slope) and the atomic component (flatter).

\begin{figure}\begin{center}
\includegraphics[clip=,width= .49\textwidth]{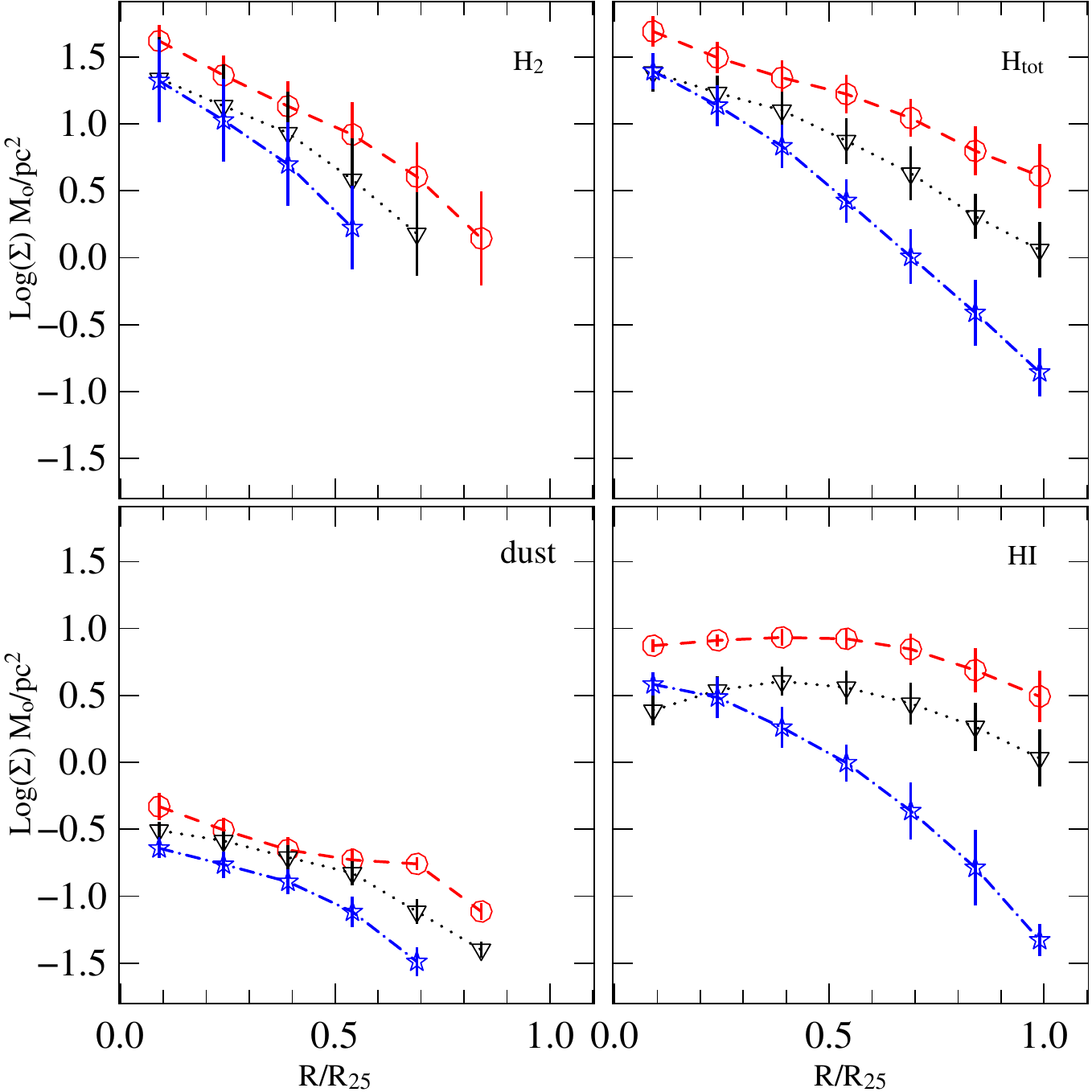}   
\end{center}\caption{Average radial surface mass densities of the dust (bottom left panel), $\Sigma_{\rm HI}$ (bottom right panel), $\Sigma_{H_2}$ (top left panel), and $\Sigma_{\rm H_{tot}}$ (top right panel). Red circles, black triangles, and blue stars represent the average profiles for non-deficient (\defhi \ $<$ 0.4), slightly HI-deficient (0.4 $\le$ \defhi \ $\le$ 0.7), and HI-deficient (\defhi $>$ 0.7) galaxies, respectively. The molecular component has been estimated using the Galactic \xco.}\label{sames}\end{figure}

To characterize the environmental effects in our sample, we show in Fig. \ref{sames} the bin-averaged profiles of the dust, atomic, molecular, and total gas surface mass densities for the three subsamples defined above. In each panel red circles, black triangles, and blue stars, indicate the non-deficient, slightly HI-deficient, and HI-deficient subsamples. For the molecular component we show only the case for the Galactic \xco, since the conclusions about the environmental effects are similar even if the \xco \ defined in Eq. \ref{scaleR} is used.

Galaxies with higher HI deficiency have steeper gas profiles and lower surface mass densities, indicating that the lack of atomic hydrogen affects the formation of the molecular gas, and thus the total gas component. However this trend is not completely obvious because, 
as shown by \citet{fum2}, HI deficiency is a condition necessary but not sufficient for molecular gas deficiency.
In our subsample of HI-deficient galaxies, according to the definition of H$_2$-deficiency of \citet{fum2}, only NGC\,4579 is strongly H$_2$-deficient, and for this reason the trend observed for the atomic hydrogen in the bottom right panel of Fig. \ref{sames} is less evident for the molecular gas component (top left panel in Fig. \ref{sames}).

As the HI deficiency increases, the surface mass densities are lower for all the ISM components. However, for the atomic hydrogen in the central regions of the disks of HI-deficient galaxies, the surface mass density appears somewhat higher than that of slightly HI-deficient galaxies.  This effect could be understood as ram-pressure-induced compression of the gas in the inner region of the disk \citep{ton} that increases locally the gas density \citep{byr}. Alternatively, the differences in surface mass density can be simply due to uncertainties in the inclination corrections for two of the HI-deficient galaxies, NGC\,4388 and NGC\,4402, which are nearly edge-on.

Dust radial profiles of HI-deficient galaxies (blue stars) decrease steeply at $R/R_{25} >$ 0.4, with respect to the non-deficient galaxies (red circles), but have comparable radial slopes in the inner parts of the disk, since the dust is more confined to the disk where ram pressure is less efficient in the ISM removal. Such a trend has already been found by \citet{cor}.

\subsection{Dust-to-gas ratios}

\begin{figure*}\begin{center}
\includegraphics[clip=,width= .99\textwidth]{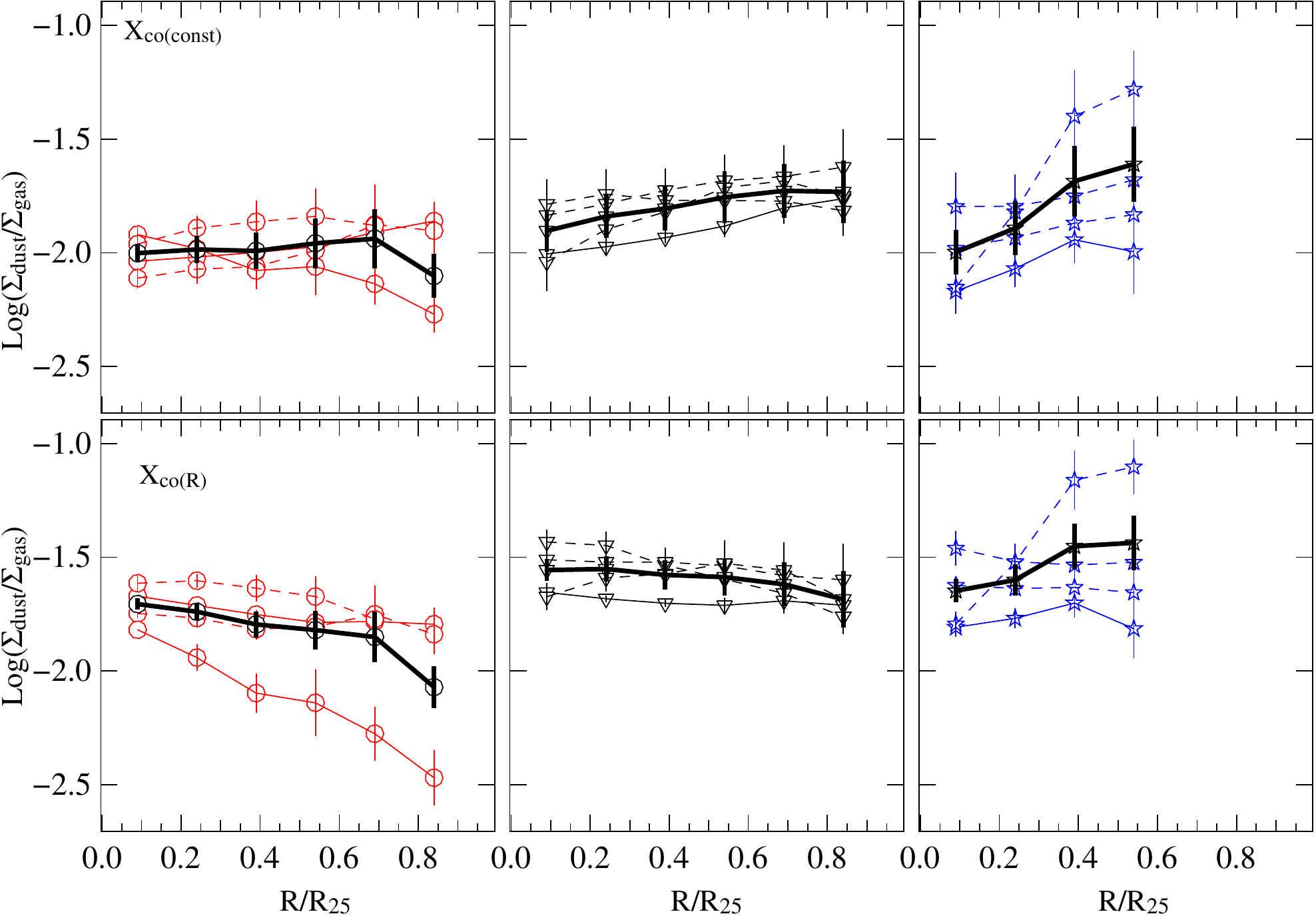}   
\end{center}\caption{Dust-to-gas ratio as a function of $R/R_{25}$ for non-deficient (red circles), \defhi $<$ 0.4, slightly HI-deficient 0.4 $\le$ \defhi $\le$ 0.7, and HI-deficient galaxies (blue stars), \defhi $>$ 0.7. The solid line shows the profile for the new mapped galaxies. The black solid line shows the average profiles. The molecular content has been estimated using both the Galactic \xco\ (top row) and the \xco\ of Eq. \ref{scaleR} (bottom row).}\label{dtgsample}\end{figure*}

In Fig. \ref{dtgsample} we show the radial profiles of the dust-to-gas ratio for the three subsamples of galaxies, obtained after convolving the gas components to the resolution of the dust mass maps. The black solid line in each panel shows the bin-averaged profiles for each subset.

In the top panels of Fig. \ref{dtgsample} we use the constant Galactic \xco. The averaged profiles show only moderate change with HI deficiency; the dust-to-gas ratio is almost constant throughout the disk for non-deficient galaxies (top left panel of Fig. \ref{dtgsample}), while it increases with radius for the strongly HI-deficient subset (top right panel of Fig. \ref{dtgsample}). This could be explained because the gas, particularly the atomic component, is more easily stripped with respect to the dust \citep{cor2}, thus producing a rapid increase of the dust-to-gas ratio at R/R$_{25} >$ 0.2. However, the average slope of this subset may be biased by the highly HI-deficient galaxy NGC\,4569 (\defhi = 1.47), which has a gas profile that decreases rapidly at small radii.

Similar trends with HI deficiency are found when the metallicity dependent \xco\ is used (bottom panels of Fig. \ref{dtgsample}). However, there is a systematic change in the dust-to-gas ratio. Because of the lower \xco\ in the metal-rich nuclei, the H$_2$-dominated total gas mass surface densities decrease. Thus, the central dust-to-gas ratio increases, consequently producing negative radial gradients. As shown in the bottom left panel of Fig. \ref{dtgsample}, this effect is more evident for non-deficient objects, confirming what was already found by \citet{mag} and by \citet{foy} on the nearby galaxy M\,83. Also, \citet{mun}, using Spitzer data, find decreasing dust-to-gas profiles for an \xco \ factor that depends on the metallicity gradient. Simple considerations imply that the dust-to-gas ratio should follow linearly the metallicity; both dust-to-gas and metallicity trace metals, the former in the solid and the latter in the gaseous phase \citep{dra2}. Thus, a lower-than-Galactic \xco \ in a galaxy's center appears to be necessary for the dust-to-gas ratio to have the same gradient observed for the metallicity. However, because of the large uncertainties in the absolute calibration of the metallicity indicators \citep[e.g.][]{mag}, this statement is inconclusive. Indeed, independently of the metallicity, when the surface mass density is dominated by the atomic gas at all radii the dust-to-gas ratio decreases with radius, similarly to the metallicity gradient (see e.g. \citealt{ben2} for NGC 2403, and \citealt{fri}; \citealt{smi3} for the Andromeda galaxy).

\begin{table*}
\caption{Dust and gas properties in the sample}
\label{dtgvalue}\begin{center}
\begin{tabular}{c c c c c c c c c}
\hline \hline
\\
Galaxy 			& M$_{dust}$ 		& M$_{HI}$ 		& M$_{H_2}$&M$_{dust}$/(M$_{HI}$+M$_{H_2}$)	&f$_{mol}$\\				&10$^7$ M$_\odot$	&10$^8$ M$_\odot$	&10$^8$ M$_\odot$		&		&		\\
(1) 			& (2)			&(3)			&(4) 				& (5) 				& (6)					\\
\hline
NGC 4189 		& 4.47$\pm$ 0.78 	& 22.7$\pm$0.7 	& 17.4$\pm$1.34 		&0.011					&0.43\\
NGC 4294 		& 1.15$\pm$0.23 	&18.4$\pm$1		& 0.73$\pm$0.2		&0.006					&0.04\\
NGC 4298 		& 2.29$\pm$0.4 	&5.6$\pm$0.9		& 10.4$\pm$0.8		&0.014					&0.65\\
NGC 4299 		& 0.52$\pm$0.1	&12.4$\pm$0.5		& 1.1$\pm$0.2			&0.004					&0.08\\
NGC 4351 		& 0.4$\pm$0.07 	&3.4$\pm$0.4		& 0.71$\pm$0.2		&0.01					&0.17\\
NGC 4388 		& 1.2$\pm$0.24 	&4.15$\pm$2		& 7.13$\pm$0.6		&0.01					&0.63\\
NGC 4424 		& 0.59$\pm$0.1	&3.97$\pm$0.2		& 4.2$\pm$1.2	  		&0.007					&0.51\\
\hline \hline\end{tabular}\end{center} 
Column 1: galaxy name; col. 2: dust mass \citep{dav}; col. 3: integrated HI mass \citep{chu}; col. 4: integrated H$_2$ mass (this work); col. 5: dust-to-gas ratio; col. 6: molecular-to-total gas mass fraction. The molecular masses have been estimated using the Galactic \xco.
\end{table*}

\section{The linear relation between $\Sigma_{gas}$ and F$_{250}$: a pixel by pixel analysis}
\label{pixpix}

\begin{figure}[!h]
\begin{center}
\includegraphics[angle=90,clip=,width= .38\textwidth]{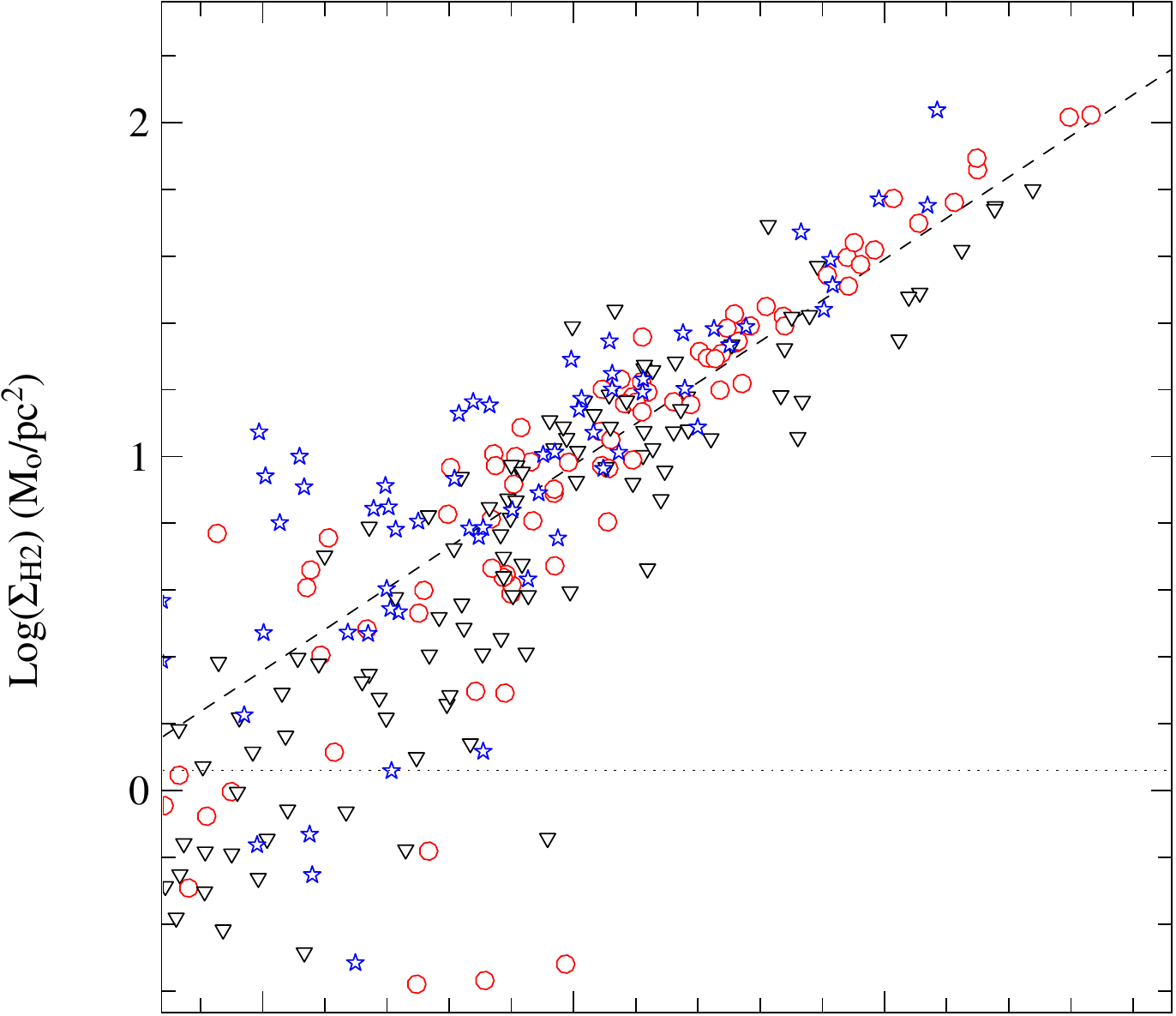}
\includegraphics[angle=90,clip=,width= .38\textwidth]{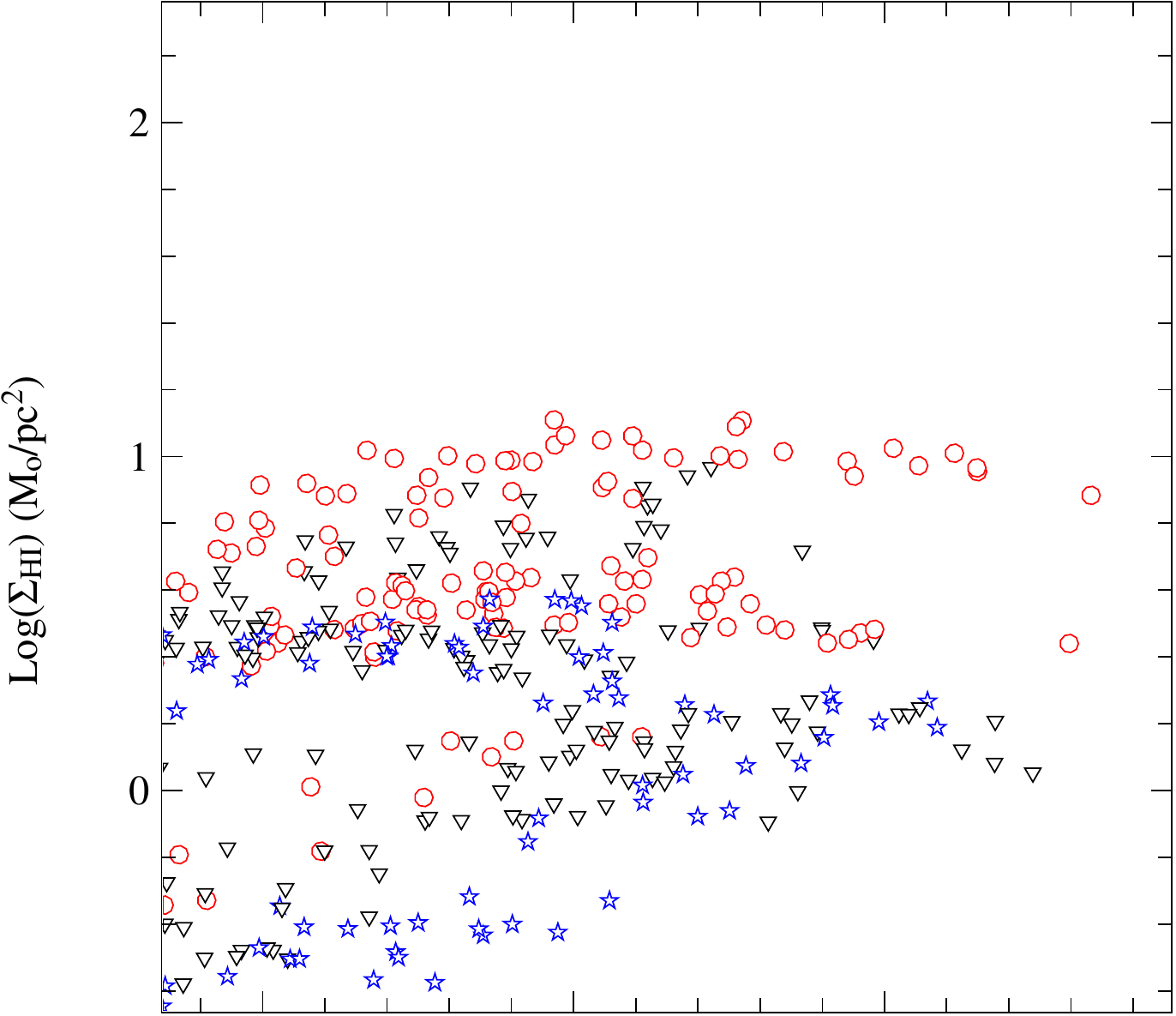}
\includegraphics[clip=,width= .38\textwidth]{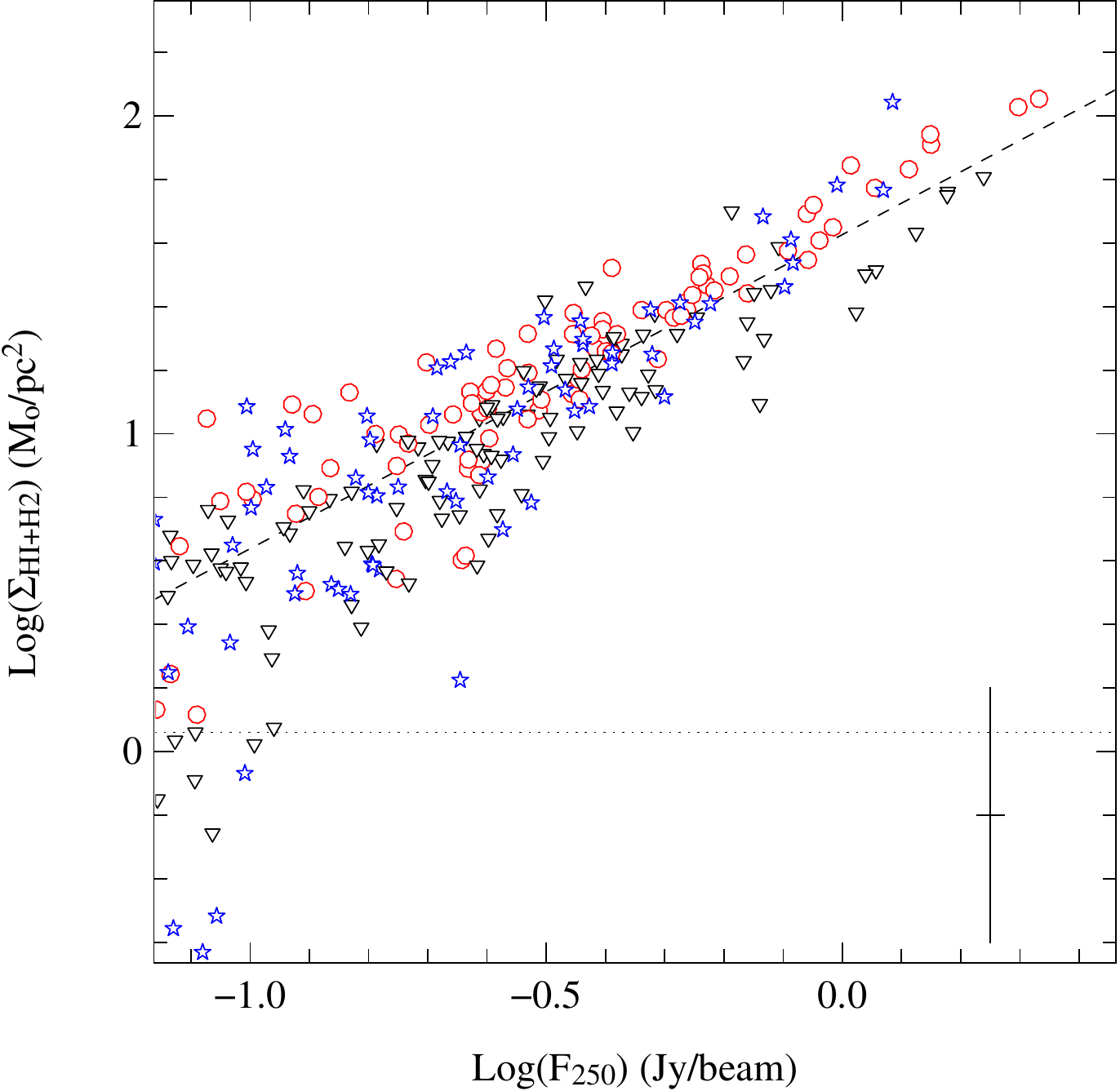}
\end{center}
\caption{Surface mass densities of molecular hydrogen (top panel), atomic hydrogen (middle panel), and total gas (bottom panel) as a function of the surface brightness at 250 $\mu$m on a pixel-by-pixel basis. Different symbols distinguish non-deficient (red circles), slightly HI-deficient (black triangles), and HI-deficient galaxies (blue stars). The pixel size has been set to 36$''$. The horizontal dotted lines show the 3$\sigma$ level for the molecular and total gas (3$\sigma$ = 1.15 M$_\odot$ pc$^{-2}$). The diagonal lines show the best linear fits. $\Sigma_{H_2}$ in the top and bottom panels has been estimated using the constant Galactic \xco. A typical error bar is shown in the bottom panel.}
\label{fcosample}\end{figure}

Fig. \ref{fcosample} shows the surface mass density of H$_2$ (top panel, using a Galactic \xco), HI (middle panel), and the total gas (bottom panel) as a function of the surface brightness at 250 $\mu$m images ($F_{250}$) on a pixel-by-pixel basis. We chose this band because of its good sensitivity and better spatial resolution. Moreover, the surface brightness at one wavelength is preferred over the dust surface mass density because it is related to directly observable quantities; the scatter in the relation increases if the dust surface mass density is used instead of the 250 $\mu$m surface brightness. Because of the poorer resolution of some of the HI maps, we smoothed all the data to a beam size of 36$''$, and regridded all the maps to the same pixel size. Each pixel covers an area of $\sim$ 9 kpc$^2$, except for NGC 4189 (at a distance of 32 Mpc), for which the area corresponds to $\sim$ 30 kpc$^2$. The top panel of Fig. \ref{fcosample} shows that $\Sigma_{H_2}$ is proportional to the surface brightness at 250 $\mu$m. Considering all pixels above the 250 $\mu$m surface brightness 3$\sigma$ threshold value of 0.07 Jy beam$^{-1}$ and $\Sigma_{H_2} \geq 3 \sigma$ = 1.15 M$_\odot$ pc$^{-2}$, we found a best linear fit:

\begin{equation}
log \ \Sigma_{H_2} = 1.23\pm0.05 \cdot log \ F_{250} + 1.59\pm0.03,
\label{h2rel}
\end{equation} 
\noindent
with a Pearson correlation coefficient $r=0.86$. 

The relation is tighter for non-deficient galaxies (red circles) with respect to galaxies with a pronounced HI-deficiency (blue stars). When computing a best linear fit we found a slope $m = 1.29\pm0.07$ ($r$ = 0.91) for the non-deficient subsample, and $m = 1.08\pm0.1$ ($r$ = 0.81) for the HI-deficient one. For both subsamples we find a similar intercept, $\sim$ 1.6. The slightly better correlation between the 250 $\mu$m surface brightness and $\Sigma_{H_2}$ for non-deficient objects could indicate that the environment perturbs both the molecular and dust disk, but in somewhat different ways, thus increasing the scatter. 

In the middle panel of Fig. \ref{fcosample} no clear correlation can be seen between $\Sigma_{HI}$ and $F_{250}$. The spread is larger than for $\Sigma_{H_2}$ and it appears that non-deficient galaxies have higher $\Sigma_{HI}$ ($\sim$ 5 M$_\odot$ pc$^{-2}$ for $F_{250} \sim$ 0.3 Jy beam$^{-1}$) than HI-deficient ones ($\sim$ 0.5 M$_\odot$ pc$^{-2}$), though both samples contain outliers. This indicates that uncertainties in the \defhi\ parameter prevent a clear characterization of the environment effects on all galactic properties.

Adding the HI contribution to all pixels with detected molecular gas, as shown in the bottom panel of Fig. \ref{fcosample}, we find:
\begin{equation}
log \ \Sigma_{gas} = 0.99\pm0.04 \cdot log \ F_{250} + 1.63\pm0.02.
\label{eales2}
\end{equation} 
\noindent
The slope in this case is shallower than that in Eq. \ref{h2rel} and roughly linear. However, the goodness of the correlation does not change significantly when adding HI ($r$ = 0.87). This is in contrast with the correlation found by \citet{corb2} between the integrated properties: the total gas mass correlates better globally with the dust mass than the single gas components. A similar discrepancy is seen in the relations between star formation rate and gas mass surface densities; while a strong correlation is found between the {\it global} star formation rates and the total gas mass surface densities \citep{kenn3}, a {\it resolved} analysis gives a better correlation with the molecular gas surface density \citep{big,sch}.

If we use an \xco\ that varies radially according to Eq. \ref{scaleR}, the correlation between $\Sigma_{gas}$ and $F_{250}$ is slightly degraded ($r$ = 0.8), and the slope becomes shallower ($m$ = 0.75$\pm$0.04). The radial variation of the \xco\ produces in fact flatter radial profiles for the molecular surface mass density, which explains the shallower slope.

\citet{eal2} proposed that dust emission can be used to map the ISM in high redshift galaxies: adopting the emissivity of Galactic dust, they derived a linear correlation between the monochromatic luminosity at 250 $\mu$m and the total mass of hydrogen (using the Galactic \xco\ and calibrating the relation on a few nearby galaxies observed with {\it Herschel} and with gas mass estimates). With the dust temperatures measured on our maps \citep[$T \sim $ 20 K, see][]{dav2}, their correlation can be converted to $log \ \Sigma_{gas} = log \ F_{250} + 1.62$, in good agreement with Eq. \ref{eales2}. However, we had to assume that the 250 $\mu$m luminosity arises from the same region occupied by gas, which is guaranteed for the molecular gas component only.

\section{Conclusions}
\label{conc}

In this paper we present new IRAM-30m \couno \ and \codue \ observations of a sample of spiral galaxies in the Virgo cluster which are in the HeViCS field and in the VIVA sample. We observed four galaxies in OTF mode and three in PS mode. For observations done in OTF mode CO-maps are discussed; for galaxies observed in PS mode we only estimate the total molecular content. We include in our analysis a sample of eight galaxies observed in \couno \ by \citet{kun}, and investigate the effects of the environment on the surface mass density of the atomic and molecular gas phases and on the dust. Maps of the dust surface mass density have been estimated using the  HeViCS data set \citep{dav,aul}. Finally we analyze the relation between the different gas components and the dust distributions in relation to the \defhi\ parameter. Our main conclusions are the following:

\begin{enumerate}
\item For the new observed galaxies the global molecular gas-to-total mass fraction in the sample varies in the range 0.04 $\le f_{mol} \le$ 0.65, using the Galactic \xco, with the undisturbed galaxies showing the lowest values. Disturbed galaxies have a lower atomic and molecular content but the environment is more efficient at removing the HI than the molecular gas.\item The total gas surface mass density in the central regions of the disk, above a threshold of 9 M$_\odot$ pc$^{-2}$ is preferentially found in molecular form, in agreement with the results of \citet{big}. However, a few HI-deficient galaxies show an order of magnitude lower threshold, indicating that the value may depend on local conditions. Highly HI-deficient galaxies have H$_2$, HI, and dust surface mass densities in the the central region lower by about a factor 2, suggesting gas and dust removal even in the inner disks. 
\item Galaxies with high HI deficiency have steeper radial profiles of the total gas, as a consequence of the combined effect of HI and H$_2$ deficiency, in agreement with previous studies \citep{fum2,cor2}. The average radial profiles of the dust surface mass density in HI-deficient galaxies is steeper than in HI-normal galaxies. This is consistent with the results found by \citet{cor}, on the radial distribution of the flux density at 250, 350 and 500 $\mu$m.
\item The dust-to-gas ratio for non-deficient galaxies has a radial trend which depends on the \xco \ conversion factor, as already discussed by \citet{mag}. This ratio decreases radially only if \xco \ increases radially due to metallicity gradients. For HI-deficient galaxies instead, the dust-to-gas ratio stays constant or increases radially, independently of the \xco \ factors used. This indicates that atomic gas is stripped more efficiently than the dust in a cluster environment. Since both dust and gas radial profiles truncate sharply well inside the optical disk, the dust-to-gas ratio cannot be traced at large galactocentric radius in highly perturbed galaxies.
\item  We observe a tight pixel-by-pixel correlation between the mass surface density of the molecular hydrogen and the 250 $\mu$m surface brightness. In HI-deficient galaxies the correlation weakens, suggesting that the environment is perturbing both gas and dust. Adding the atomic gas to the molecular component makes the correlation with the 250 $\mu$m surface brightness linear, but the correlation coefficient does not change significantly. Thus it is not clear if the dust emission at long FIR wavelengths can be used as a tracer of the total gas surface mass density, or of the molecular gas only.
\end{enumerate}

\begin{acknowledgements}
We thank the referee J. Kenney for his useful comments which lead to substantial improvement of the original manuscript.
We would like to thank D. Munro for freely distributing his Yorick programming language (available at \texttt{http://www.maumae.net/yorick/doc/index.html}).
This work has benefited from research funding from the European Community's sixth Framework Programme (RadioNet).
C. P., L. M., S. B., L. H., E. C., C.G., S. di S. A., are supported through the ASI-INAF agreement I/016/07/0 and I/009/10/0.
 L. H. and C. P. are also supported by a PRIN-INAF 2009/11 grant.
 The research leading to these results has received funding from the European Community's Seventh Framework Programme (/FP7/2007-2013/) under grant agreement No 229517. C. P. would like to thank Fabrizio Massi, and Riccardo Cesaroni for their help during data reduction. C. P. also thanks Christof Buchbender for the really productive "tapas y cerveza" night in Granada.
 \end{acknowledgements}
   
\appendix
\section{Notes on individual objects}
\label{single}

\subsection{NGC 4189}

NGC 4189 is a relatively undisturbed Sc late-type galaxy (\defhi \ = 0.25) located at the edge of the Virgo cluster. The determination of its distance and the membership to the cluster is debated, because Virgo is a cluster that is still forming and comprises distinct clouds of galaxies at different distances. Using the relation of \citet{tul} in B-band, \citet{yas} estimated for NGC 4189 a distance of 33.8 Mpc, i.e. more than twice the distance of the Milky Way from the center of the cluster ($\sim$ 17 Mpc). \citet{sol}, starting from the HI distribution, reconstructed the structure of the cluster, and found galaxies with elongated gas-tails, due to environmental effects, out to cluster centre distances of 25-30 Mpc. These results imply that at the distance of NGC 4189 from the cluster center it is still possible to find galaxies that are Virgo members. In the following we considered for NGC 4189 the average distance of 32 Mpc for the sub-group to which NGC 4189 belongs, determined using the Tully-Fisher relation in H-band \citep{gav}. The HI disk is more extended than the optical disk, showing a symmetric warp \citep{chu} with enhanced emission in the South-East regions, in correspondence to a ridge of H$\alpha$ emission \citep{koo}.

\subsection{NGC 4298}

NGC 4298 is an Sc galaxy (\defhi\ = 0.41) located in the North-East region of Virgo at a projected distance of 3.2$^\circ$ from M87. The galaxy has flocculent spiral structure \citep{elm} and has a companion, NGC 4302, at a projected distance of 11 kpc, having a comparable velocity ($\Delta {\rm v} \approx$ 30 km s$^{-1}$). Both galaxies show HI-tails pointing away from M87, probably due to ram pressure stripping and/or tidal interaction. However the undisturbed stellar component of NGC 4302 favors the ram pressure hypothesis \citep[see][]{chu2}. The HI disk in NGC 4298 is compressed in the South-East region, in the direction of M87, as a consequence of the ram pressure exerted by the intracluster medium. The H$\alpha$ emission is mostly concentrated in the central region of the disk, except for an enhanced arc in South-East direction, in correspondence with the HI compressed region \citep{koo}.

\subsection{NGC 4299 \& NGC 4294}

NGC 4299 is an Scd galaxy at a distance of 2.5$^\circ$ from M87 (\defhi \ = -0.43) which forms a possible interacting pair with NGC 4294 (\defhi \ = -0.11). However both galaxies retain their HI gas which forms an extended HI disk \citep{chu} with a long HI tail to the South West. \citet{koo} showed that this pair has enhanced H$\alpha$ emission in the northern part of the disk, where the HI is compressed. This indicates a stronger star formation rate, probably due to tidal interaction.

\subsection{NGC 4351}

NGC 4351 is an Sc galaxy which is close to the cluster centre, about 1.7$^\circ$ from M87 and is slightly HI-deficient (\defhi \ = 0.23). It shows a compression in the HI emission in the north eastern region, due to ram pressure \citep{chu}. The brightest optical region is offset with respect to the center, and also the H$\alpha$ emission is asymmetric with enhanced emission in a circumnuclear region \citep{koo}. 

\subsection{NGC 4388}

NGC 4388 is an Sab galaxy located at a projected distance of 1.3$^\circ$ from M87. The galaxy is one of the most HI-deficient in our sample (\defhi\ = 1.16), showing HI distribution truncated within the optical disk \citep{chu}. \citet{yos} found that the H$\alpha$ emission extends for 35 kpc in the north east direction. \citet{vol} found HI gas out to 20 kpc along the same direction as the H$\alpha$ emission. Further observations showed that the HI tail is even more extended, 110 kpc, with a mass of 3.4 $\times$ 10$^8$ M$_{\odot}$ \citep{oos}. The most probable origin of this elongated tail is a ram pressure stripping episode that occurred about 200 Myr ago \citep{pap}.

\subsection{NGC 4424}

NGC 4424 is an Sa galaxy with \defhi = 0.97. \citet{chu2} showed a long HI tail pointing away from M87 and \citet{ken} observed star forming regions in the central kpc, but no star formation has been detected beyond these regions. \citet{cort} found two lobes in the molecular gas component in the central 3 kpc, with total mass M$_{H_2} \sim$ 4.0 $\times$ 10$^{7}$ M$_{\odot}$. The central regions are highly disturbed  and asymmetric in the radio continuum, probably as a consequence of strong gravitational interaction \citep{chu}.


\end{document}